\documentclass[fleqn]{aa}
\usepackage{times}
\usepackage{graphics}

\begin{document}


   \title{Line formation in solar granulation}

   \subtitle{VI. [C\,{\sc i}], C\,{\sc i}, CH and C$_2$ lines 
and the photospheric C abundance}

   \author{M. Asplund$^{1}$,
           N. Grevesse$^{2,3}$, A. J. Sauval$^{4}$,
           C. Allende Prieto$^{5}$ and
           R. Blomme$^{4}$
          }

   \institute{
$^{1}$ Research School of Astronomy and Astrophysics,
Mt. Stromlo Observatory, Cotter Rd., Weston,
ACT 2611, Australia\\
$^{2}$
Centre Spatial de Li\`ege, Universit\'e de Li\`ege,
avenue Pr\'e Aily, B-4031 Angleur-Li\`ege, Belgium \\
$^{3}$ Institut d'Astrophysique et de G\'eophysique, Universit\'e de Li\`ege,
All\'ee du 6 ao\^ut, 17, B5C, 4000 Li\`ege, Belgium \\
$^{4}$ Observatoire Royal de Belgique, avenue circulaire, 3,
B-1180 Bruxelles, Belgium \\
$^{5}$ McDonald Observatory and Department of Astronomy,
University of Texas, Austin, TX 78712-1083, USA \\
             }

   \offprints{\email{martin@mso.anu.edu.au}}

   \date{Received: yes; accepted: yes}

\abstract{
The solar photospheric 
carbon abundance has been determined from [C\,{\sc i}], C\,{\sc i}, 
CH vibration-rotation, CH A-X electronic and 
C$_2$ Swan electronic lines by means of
a time-dependent, 3D, hydrodynamical model of the solar atmosphere.
Departures from LTE have
been considered for the  C\,{\sc i} lines. These turned out to
be of increasing importance for stronger lines and are crucial to
remove a trend in LTE abundances with the strengths of the lines. 
Very gratifying agreement is found among all the atomic and
molecular abundance diagnostics in spite of their widely different
line formation sensitivities.
The mean of the solar carbon abundance
based on the four primary abundance indicators 
([C\,{\sc i}], C\,{\sc i}, CH vibration-rotation, C$_2$ Swan) is 
${\rm log} \, \epsilon_{\rm C} = 8.39 \pm 0.05$,
including our best estimate of possible systematic errors. 
Consistent results also come from the CH electronic lines, which
we have relegated to a supporting role due to their sensitivity
to the line broadening.  
The new 3D based solar C abundance is significantly lower than previously 
estimated in studies using 1D model atmospheres.
}

\authorrunning{M. Asplund et al.}
\titlerunning{Solar line formation: VI. The photospheric C abundance}

   \maketitle

\section{Introduction}

Carbon is a crucial element for the emergence of life in the Universe
as we know it. Furthermore, it is one of the most abundant elements
in the cosmos behind only hydrogen, helium and oxygen in this respect. While
it is clear that carbon is produced through the $3\alpha$-reaction
(Burbidge et al. 1957)
in stellar interiors, a debate regarding the exact site of
its production is still ongoing.
Recently both low- and intermediate mass stars as well as
massive stars have been advocated to be the primary sources
of the element (e.g. Chiappini et al. 2003; Akerman et al. 2004).
Carbon also plays a key role in the physics of the interstellar
medium for example 
through its propensity to form dust (e.g. Dopita \& Sutherland 2002;
Lodders 2003).

In view of its rather special status in astrophysics, it is clearly
of particular importance to ascertain the traditional abundance anchorage
point, namely the solar carbon abundance. In the standard reference
of Anders \& Grevesse (1989) the solar carbon abundance is quoted as
${\rm log} \, \epsilon_{\rm C} = 8.56 \pm 0.04$, which was subsequently
revised to ${\rm log} \, \epsilon_{\rm C} = 8.52 \pm 0.06$ by
Grevesse \& Sauval (1998) based on a re-analysis of various atomic
and molecular lines using a temperature-modified Holweger-M\"uller (1974) 1D
semi-empirical model atmosphere. Holweger (2001) preferred the slightly higher
value of ${\rm log} \, \epsilon_{\rm C} = 8.59 \pm 0.11$ based only
on permitted C\,{\sc i} lines in the optical and infrared (IR).
In sharp contrast to these results, Allende Prieto et al. (2002) found
the much lower value ${\rm log} \, \epsilon_{\rm C} = 8.39 \pm 0.04$
through a detailed study of the forbidden [C\,{\sc i}]
line at 872.7\,nm employing a state-of-the-art 3D time-dependent
hydrodynamical model of the solar photosphere (Asplund et al. 2000b).
This latter value is in good agreement with the evidence from
nearby B stars (Sofia \& Meyer 2001), local interstellar medium
(Meyer et al. 1998; Andr\'e et al. 2003), the solar corona and the solar wind
(Murphy et al. 1997; Reames 1999), in particular in view
of the recent downward revision of the solar oxygen abundance
(Allende Prieto et al. 2001; Asplund et al. 2004, hereafter Paper IV).

While the [C\,{\sc i}] 872.7\,nm line has a well-determined
transition probability and should be robust against
departures from LTE, its relative weakness and
the possible presence of weak blends
may cast some doubt on the reliability of the low carbon
abundance derived by Allende Prieto et al. (2002).
We here present a comprehensive study of other available
carbon diagnostics, including permitted atomic and
various molecular lines using the same 3D hydrodynamical solar
model atmosphere.
Our results corroborate the findings of Allende Prieto et al.
and unambiguously point to a significant revision by $-0.17$\,dex
of the solar carbon abundance compared with the often-adopted
value given in Anders \& Grevesse (1989).
It is noteworthy that only with a 3D analysis are the primary
atomic and molecular abundance criteria in excellent agreement,
which further strengthens our conclusions.

\section{Analysis}

\subsection{Atomic and molecular data
\label{s:atomicdata}}

\noindent
{\bf [C\,{\sc i}] line:}
Recent calculations of the transition probability of the [C\,{\sc i}] 872.7\,nm line
have apparently converged to $A = 0.640$\,s$^{-1}$ or log\,$gf = -8.136$
(Hibbert et al. 1993; Galavis et al. 1997).
The central wavelength of the line is used as a free parameter in the $\chi^2$ analysis
of the profile fitting described below, which results in a rest wavelength
of $872.7139 \pm 0.0004$\,nm when taking into account the predicted convective
blue-shift of the line.
The lower level excitation potential of the line is $1.264$\,eV
(Bashkin \& Stoner 1975).

\noindent
{\bf C\,{\sc i} lines:}
The transition probabilities for the selected 16 permitted C\,{\sc i} lines 
which represent our primary line list
were taken from Hibbert et al. (1993).
The VALD\footnote{http://www.astro.uu.se/$\sim$vald}
(Piskunov et al. 1995; Kupka et al. 1999) and 
NIST\footnote{http://physics.nist.gov/cgi-bin/AtData/main\_asd} 
(Wiese et al. 1996)
databases provided the necessary data for excitation potential,
radiative broadening and central wavelengths.
Collisional broadening by H
was computed according to the tabulations in
Anstee \& O'Mara (1995), Barklem \& O'Mara (1997) and Barklem et al. (1998), 
except for C\,{\sc i} 505.2, 538.0, 658.7, 3085.4 and 3406.5\,nm which fall outside 
the table boundaries. For those lines the classical Uns\"old (1955) 
recipe with an enhancement factor $E=2.0$ was used. The corresponding uncertainty
is however minor: with $E=1.0$ the derived abundances
for the three optical lines would be about 0.01\,dex higher while
for the two IR lines the corresponding error is about 0.03\,dex.

We initially considered additional C\,{\sc i} lines, which however were
all rejected in the final analysis for various reasons, such as
suffering from blends (e.g. 477.5, 1186.3\,nm), too large sensitivity to
velocity and/or pressure broadening (e.g. 965.8, 1068.5, 1070.7, 1189.5,
1744.8, 3372.8\,nm) or uncertain continuum placement (e.g. 1184.8\,nm).
We stress that it is better to retain only the best quality abundance
indicators than using a larger sample of less reliable lines.
Nevertheless, our line list is sufficiently large to secure a final
mean abundance that is insensitive to possible problems with a few individual lines.

\noindent
{\bf CH vibration-rotation and electronic lines:}
The analysis is based on the best 102 vibration-rotation lines of the (1,0), (2,1) and
(3,2) bands in the infrared ($3.3-3.8\,\mu$m). We
also include, with a lower weight, nine apparently clean weak
lines from the (0,0) and (1,1) bands of CH A-X 
around 430\,nm (6 and 3 lines, respectively).
The dissociation energy of CH is well-known:
$D_0 {\rm (CH)} = 3.465$\,eV (Huber \& Herzberg 1979). 
The partition functions and statistical
weights for CH are
taken from Sauval \& Tatum (1984) for consistency 
with the adopted dissociation energy and
equilibrium constants. The lower level excitation potentials for the CH lines were
adopted from M\'elen et al. (1989) and the transition probabilities from 
Follmeg et al. (1987).
The latter are the same as used by Grevesse et al. (1991).
The band oscillator strengths for the CH A-X lines are
$f_{00} = (5.16 \pm 0.11) \cdot 10^{-3}$ and
$f_{11} = (4.46 \pm 0.15) \cdot 10^{-3}$ (Larsson \& Siegbahn 1986).
As the quantum mechanical approach for pressure broadening
outlined by O'Mara and collaborators (e.g. Anstee \& O'Mara 1995) is not
applicable to molecular transitions, we have resorted to the classical
Uns\"old (1955) recipe with an enhancement factor of 2 for all molecular lines.
The particular choice of enhancement factor has no impact on 
the derived abundances however for these lines.

\noindent
{\bf C$_2$ electronic lines:}
We selected the 17 least blended weak C$_2$ lines 
from the (0,0) $d^3 \Pi_g - a^3 \Pi_u$ Swan band.
Our adopted dissociation energy of C$_2$ comes from the accurate experimental
measurement by Urdahl et al. (1991):
$D_0 {\rm (C_2)} = 6.297$\,eV. This laboratory determination is in very good
agreement with the theoretical calculations of Pradham et al. (1994) which
suggests $D_0 {\rm (C_2)} = 6.27$\,eV. 
Our value is 0.09\,eV higher than that recommended
by Huber \& Herzberg (1979) and 0.19\,eV higher than the value employed by
Lambert (1978) in his classical solar abundance analysis of CNO.
We remind the reader that the derived
carbon abundance from C$_2$ lines depends on the dissociation energy as
$\Delta {\rm log} \epsilon_{\rm C} \approx - \Delta D_0 / 2$.
The C$_2$ partition functions and equilibrium constants
have been adopted from Sauval \& Tatum (1984).
The C$_2$ Swan band oscillator strengths are well-determined from measurements
of the radiative lifetime of the $d^3 \Pi_g$ state (see Grevesse et al. 1991).
As for the CH lines, we use an enhancement factor of 2 to the
Uns\"old recipe for pressure broadening, which however has no
influence on the final results.
We do not consider the lines of the C$_2$ Phillips system to be
sufficiently reliable for our purposes and hence do not include
any such lines in the analysis.

\noindent
{\bf CO vibration-rotation lines:} In principle, CO lines can be utilised
to derive the carbon abundance whenever
the oxygen abundance has first been determined accurately from
alternative diagnostics.
We will not consider CO in this analysis, however, due to their large
model atmosphere sensitivity through the high dissociation energy which
makes them less reliable than other diagnostics.
Indeed, no available 1D hydrostatic model atmosphere, be it semi-empirical like
the Holweger-M\"uller (1974) or VAL-3C (Vernazza et al. 1976) models or
theoretical like the {\sc marcs} (Gustafsson et al. 1975 and subsequent updates)
or Kurucz (1993) models, have been able to explain the strong CO lines observed
in the Sun (not counting of course semi-empirical models specifically constructed
to reproduce the CO lines, such as the models described by 
Grevesse \& Sauval 1991, 1994, Grevesse et al. 1995 and Avrett 1995).
This has been taken as evidence for a very cool
temperature structure or evidence for a temperature bifurcation of the
gas in the upper photosphere/lower chromosphere (e.g. Ayres 2002 and references therein).
Uitenbroek (2000a,b) has shown that
the situation has improved with the advent of 1D time-dependent hydrodynamical
atmosphere models which account for important non-LTE effects
(Carlsson \& Stein 1992, 1995, 1997) and 3D hydrodynamical models like those
described herein.
Indeed, with the same 3D model as used here the long-standing COnundrum
may finally be about to be resolved (Asensio Ramos et al. 2003).
We intend to return to the CO lines in a future investigation in this series.

\subsection{Observational data}

Throughout we use disk-center ($\mu = 1.0$ for the optical lines and
$\mu = 0.935$ for molecular lines in the IR) intensity observations.
In the optical wavelength region the Brault \& Neckel (1987, see also
Neckel 1999) solar atlas has been employed. This high-quality atlas was recorded
with the Fourier Transform Spectrograph at Kitt Peak Observatory at
a resolving power of about 500,000 and a signal-to-noise exceeding 1000
at all wavelength regions of interest here. Minor adjustments to the
tabulated continuum level immediately surrounding the relevant
spectral lines were made. 
In some cases the Kitt Peak FTS atlas was supplemented with the
Jungfraujoch disk-center solar atlas (Delbouille et al. 1973)
when lines were affected by telluric lines (e.g. C\,{\sc i} 711.1 and 960.3\,nm). 
For the four lines between 1254 and 1259\,nm we used the infrared solar
atlas of Delbouille et al. (1981) also recorded at Kitt Peak.
For the analysis of the last four C\,{\sc i} lines in Table \ref{t:nlte}
and the CH vibration-rotation lines
the Spacelab-3 {\sc atmos}\footnote{
http://remus.jpl.nasa.gov/atmos/ftp.sl3.sun.html}
solar disk-center intensity IR observations has been used
(Farmer \& Norton 1989, see also Farmer 1994).
The resolving power of the {\sc atmos} atlas is about 200,000.
The $S/N$ varies with wavelength but is typically at least 400 in the 
regions of interest here. 

A major advantage with the new generation of 3D hydrodynamical model
atmospheres employed here is the in general excellent agreement between
predicted and observed line profiles without resorting to the usual
micro- and macroturbulence necessary in 1D spectrum synthesis
(e.g. Asplund et al. 2000a,b,c).
This facilitates the use of line
profile fitting for individual lines for the carbon abundance determination,
a method that we apply here to the analysis of the atomic lines.
Since the line profile fitting procedure is more ambiguous in the 1D case
due to the poorer agreement between predicted
and observed profiles, we here adopt the theoretical line intensities
from the 3D profile fitting as ``observed'' equivalent widths for the
1D spectrum synthesis (we note that these values are in excellent agreement
with those directly measured on the observed spectra). 
This enables a direct comparison of the
effects stemming from the choice of 1D or 3D model atmospheres.
Due to the sheer number of molecular lines employed in the present study,
we rely for practical reasons on the measured equivalent widths for
both the 1D and 3D abundance analyses of the molecular transitions.
The equivalent widths of the molecular lines 
have been obtained by fitting Gaussians to the observed
profiles.
Various tests have ensured that no significant error has
been introduced by the use of fitting Gaussians rather than for example
a Lorentz profile for these particular lines.

\subsection{LTE spectral line formation}

We follow the same procedure
as in previous articles in the present series (Asplund et al. 2000b,c;
Asplund 2000; Asplund et al. 2004; Asplund 2004;
hereafter Papers I, II, III, IV, V)
of studies of spectral line formation in the solar granulation.
We employ a 3D, time-dependent, hydrodynamical simulation of the
solar surface convection as a realistic model of the solar atmosphere.
The reader is referred to Stein \& Nordlund (1998) and Paper I for further
details of the construction of the 3D solar model atmosphere.
For comparison purposes, we have also performed identical calculations
employing two widely-used 1D hydrostatic models of the solar atmosphere: the
semi-empirical Holweger-M\"uller (1974) model and a theoretical,
LTE, line-blanketed {\sc marcs} model (Asplund et al. 1997).

Equipped with the 3D hydrodynamical solar atmosphere, 3D spectral
line formation calculations have been performed for [C\,{\sc i}],
C\,{\sc i}, CH, and C$_2$ lines under the simplifying assumptions of LTE.
In addition, instantaneous chemical equilibrium (ICE) has been assumed
valid for the molecule formation.
This is a potentially serious issue since in a stellar atmosphere
the time-scale for molecule formation could be longer than the
corresponding dynamical time-scale and hence the actual molecular
number densities may be out of equilibrium (e.g. Uitenbroek 2000a,b).
To our knowledge, the only published study devoted to follow the
time-dependent chemical evolution using a reaction network is
that of Asensio Ramos et al. (2003) who investigated CO in the Sun.
They concluded that ICE is an acceptable approximation below 
heights of about 700\,km. Since our CH and C$_2$ lines are formed
in much deeper layers, this assumption is unlikely to
significantly affect our derived C abundances 
(see also Asensio Ramos \& Trujillo Bueno 2003). However, we stress that
only detailed chemical evolution calculations can confirm whether
this conclusion is correct. 
While such computations are beyond the scope of the present study,
we intend to investigate this in a forthcoming publication. 
Finally we note for completeness that in principle 
it is necessary to use the non-LTE
populations of H and C when estimating the
number densities of CH and C$_2$ even under the assumption of ICE.
Fortunately, both elements are predominantly populated by
their ground states for which LTE is an excellent assumption
in the relevant parts of the solar atmosphere. 
Therefore the use of the Saha distribution when computing
the atomic partial pressures needed for estimating the molecular densities
is fully justified.
 
The spectrum synthesis employs realistic equation-of-state and
continuous opacities
(Gustafsson et al. 1975 and subsequent updates; Mihalas et al. 1988).
The temporal coverage of the part of the solar simulation which forms the basis
of the line formation computations is 50\,min of solar-time with snapshots
every 30\,s. The comparison with observations relates to disk-center
intensity profiles ($\mu = 1.0$ and $\mu = 0.935$ depending on the line). 
As the calculations self-consistently account for
the Doppler shifts arising from the convective motions, no micro- or
macroturbulence enter the 3D spectral synthesis.
In the absence of such convective line broadening, all 1D calculations have
been performed adopting a microturbulence $\xi_{\rm turb} = 1.0$\,km\,s$^{-1}$
unless noted otherwise (e.g. Holweger \& M\"uller 1974: Blackwell et al. 1995).

\subsection{Non-LTE spectral line formation}

While non-LTE effects are expected to be modest for the other
carbon abundance diagnostics, the same can not be assumed for the high-excitation
permitted C\,{\sc i} lines employed here. In order to investigate this further
we have performed detailed non-LTE calculations
based on the Holweger-M\"uller and {\sc marcs} 1D model
atmospheres. 
Full 3D non-LTE line formation computations have recently been 
carried out for Li (Asplund et al. 2003) and O (Paper IV) for
a couple of snapshots from the same solar simulation as employed here.
As described below, the requirement for a very extensive
carbon model atom to obtain realistic results currently prevents similar
calculations for this element.
Until such a 3D non-LTE study is performed we will have to rely
on the results from the corresponding 1D case, which, however, is expected
to yield sufficiently accurate results for these high-excitation lines
formed in deep atmospheric layers.
In this connection, we note the very similar non-LTE abundance corrections
obtained in 3D and 1D for O\,{\sc i} lines of similar character as
the C\,{\sc i} lines (Paper IV; Allende Prieto et al. 2004).
We intend to return to the issue of 3D non-LTE line formation for
C\,{\sc i} lines in a later study.

Our carbon model atom consists of 217 levels in total, of which 207 belong to
C\,{\sc i}, nine to C\,{\sc ii} and one represents C\,{\sc iii}.
The atom is complete up to principal quantum number $n=9$ for the singlet and
triplet systems with no fine structure of the terms accounted for.
The highest C\,{\sc i} level corresponds to an excitation potential
$\chi_{\rm exc} = 11.13$\,eV, which is only 0.13\,eV below the ionization limit.
The energy levels are taken from the experimental data of Bashkin \& Stoner (1975).
The levels are coupled radiatively through 458 bound-bound
and 216 photo-ionization transitions with the data coming from the Opacity Project
database TOPbase (Cunto et al. 1993). Excitation and ionization due to
collisions with electrons are accounted for through the impact
approximation for radiatively allowed transitions and the van Regemorter (1962)
formalism for forbidden lines adopting an oscillator strength $f=0.01$.
No inelastic hydrogen collisions are included as the available
laboratory measurements and quantum mechanical calculations
suggest that the normally employed recipe 
(Drawin 1968, see also Steenbock \& Holweger 1984)
over-estimates the collisional cross-sections by about three orders of
magnitude (Fleck et al. 1991; Belyaev et al. 1999; Barklem et al. 2003).
We do note, however, that there is some astrophysical evidence that this
is not the case for O (e.g. Allende Prieto et al. 2004). In the absence
of suggestions to the contrary for C, 
we prefer not to employ the Drawin formula 
given the uncertainties involved. 
The 1D non-LTE calculations have been performed using the statistical equilibrium
code {\sc multi} (Carlsson 1986), version 2.2.
The non-LTE abundance corrections have been interpolated from the non-LTE
and LTE equivalent widths for four different abundances
(log\,$\epsilon_{\rm C} = 8.30, 8.40, 8.50, 8.60$).
The corresponding LTE results necessary for 
estimating the non-LTE abundance corrections
were also obtained with
{\sc multi} using a model atom with all collisional cross-sections artificially
set to extremely large values to ensure full thermalization of all levels
and transitions.

\begin{figure}[t]
\resizebox{\hsize}{!}{\includegraphics{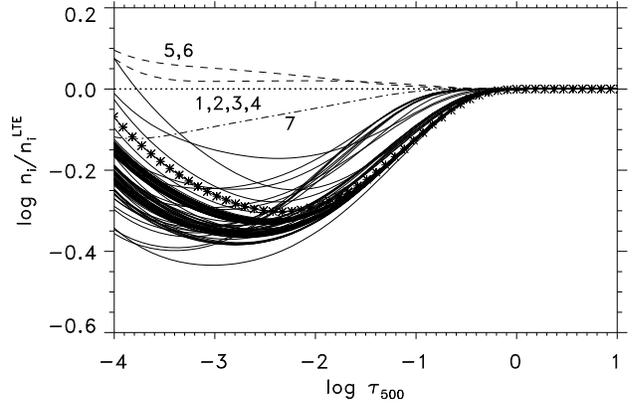}}
\caption{The departure coefficients 
$\beta_{\rm i} = n_{\rm i}/n_{\rm i}^{\rm LTE}$ for the C\,{\sc i} levels
as a function of continuum optical depth in the {\sc marcs} model atmosphere;
the corresponding case for the  Holweger-M\"uller model is almost identical.
The four lowest levels (dotted lines) show no departures from LTE, while
the following two levels (dashed lines) experience a slight over-population
in the line-forming region.
The majority of levels, however, have a small under-population (solid
lines), as does the ground state of C\,{\sc ii} (solid line connected with stars).
}
         \label{f:CI_depart}
\end{figure}

Table \ref{t:nlte} lists the resulting non-LTE abundance corrections for
the Holweger-M\"uller and {\sc marcs} 1D model atmospheres.
Clearly, the non-LTE effects are not particularly dependent on the employed model
atmosphere, at least not within the framework of 1D models. 
While the four lowest C\,{\sc i} levels
(2p$^2$ $^3$P, 2p$^2$ $^1$D, 2p$^2$ $^1$S, 2p$^3$ $^5$S$^{\rm o}$) are all
perfectly described by the Boltzmann distribution due to close
collisional coupling, the next two levels 
(3s $^3$P$^{\rm o}$, 3s $^1$P$^{\rm o}$) 
are slightly over-populated relative to the LTE expectation,
as seen in Fig. \ref{f:CI_depart}. The vast majority of levels,
however, show a very similar behaviour with small but significant underpopulations
in the typical line-forming regions
($-1 \la \log \tau_{500} \la 0$ according to the line depression contribution
functions introduced by Magain 1986):
$\beta_{\rm i} \equiv n_{\rm i}/n_{\rm i}^{\rm LTE} < 1$.
The similarity of the levels with $\chi_{\rm exc} > 8.5$\,eV
is mainly due to efficient collisional interlocking.
In all cases, the non-LTE abundance corrections for the here employed
C\,{\sc i} lines are negative
(i.e. the lines become stronger in non-LTE than in LTE),
since the line source function exceeds the Planck function
$S_{\rm l}/B_\nu \approx \beta_{\rm up}/\beta_{\rm low} < 1$ in the line-forming regions.
In addition, many of the C\,{\sc i} lines originate from
the 3s $^3$P$^{\rm o}$ or 3s $^1$P$^{\rm o}$ levels which are slightly over-populated
($\beta_{\rm low} > 1$) which increases the line opacity and hence strengthens the
lines further.
The abundance corrections show a clear
dependence on equivalent width with the largest corrections obtained for
the strongest lines.
Our non-LTE abundance corrections would have been
significantly smaller ($<0.05$\,dex)
had we included H collisions through the classical Drawin (1968) formula.
Our non-LTE results are qualitatively the same as those obtained by
St\"urenbock \& Holweger (1990) but show in general slightly larger
non-LTE effects, mainly due to our larger atom,
more realistic photo-ionization cross-sections and no inclusion of
efficient H collisions.

\begin{table}[t!]
\caption{The computed non-LTE abundance corrections  of C\,{\sc i} lines for the
Holweger-M\"uller (1974) model and {\sc marcs} (Asplund et al. 1997)
1D model atmospheres. The departures from LTE have
been estimated using {\sc multi} (Carlsson 1986) and a 217-level model atom.
\label{t:nlte}
}
\begin{tabular}{lccccc}
 \hline
line   &   	&	$\chi_{\rm exc}$ & log\,$gf$ &
\multicolumn{2}{c}{$\Delta({\rm log} \epsilon_{\rm C})$} \\
\cline{5-6}
$$[nm] & 	&	[eV]             &           &   HM & {\sc marcs} \\
 \hline
 505.2167 &  3s$^1$P$^{\rm o}_1$ -- 4p$^1$D$_2$ &  7.685 &   -1.304 &  -0.05 &  -0.04 \\
 538.0337 &  3s$^1$P$^{\rm o}_1$ -- 4p$^1$P$_1$ &  7.685 &   -1.615 &  -0.04 &  -0.03 \\
 658.7610 &  3p$^1$P$_1$ -- 4d$^1$P$^{\rm o}_1$ &  8.537 &   -1.021 &  -0.03 &  -0.03 \\
 711.1469 &  3p$^3$D$_1$ -- 4d$^3$F$^{\rm o}_2$ &  8.640 &   -1.074 &  -0.04 &  -0.04 \\
 711.3179 &  3p$^3$D$_3$ -- 4d$^3$F$^{\rm o}_4$ &  8.647 &   -0.762 &  -0.05 &  -0.05 \\
 960.3036 &  3s$^3$P$^{\rm o}_0$ -- 3p$^3$S$_1$ &  7.480 &   -0.895 &  -0.15 &  -0.15 \\
1075.3976 &  3s$^3$P$^{\rm o}_2$ -- 3p$^3$D$_1$ &  7.488 &   -1.598 &  -0.11 &  -0.10 \\
1177.7546 &  3p$^3$D$_2$ -- 3d$^3$F$^{\rm o}_2$ &  8.643 &   -0.490 &  -0.11 &  -0.09 \\
1254.9493 &  3p$^3$P$_0$ -- 3d$^3$P$^{\rm o}_1$ &  8.847 &   -0.545 &  -0.09 &  -0.08 \\
1256.2124 &  3p$^3$P$_1$ -- 3d$^3$P$^{\rm o}_0$ &  8.848 &   -0.504 &  -0.09 &  -0.08 \\
1256.9042 &  3p$^3$P$_1$ -- 3d$^3$P$^{\rm o}_1$ &  8.848 &   -0.586 &  -0.09 &  -0.08 \\
1258.1585 &  3p$^3$P$_1$ -- 3d$^3$P$^{\rm o}_2$ &  8.848 &   -0.509 &  -0.09 &  -0.08 \\
2102.3151 &  3p$^1$S$_0$ -- 3d$^1$P$^{\rm o}_1$ &  9.172 &   -0.437 &  -0.08 &  -0.06 \\
2290.6565 &  3p$^1$S$_0$ -- 4s$^1$P$^{\rm o}_1$ &  9.172 &   -0.182 &  -0.10 &  -0.08 \\
3085.4621 &  3d$^1$F$^{\rm o}_3$ -- 4p$^1$D$_2$ &  9.736 &    0.086 &  -0.05 &  -0.05 \\
3406.5790 &  4p$^1$P$_1$ -- 4d$^1$D$^{\rm o}_2$ &  9.989 &    0.454 &  -0.11 &  -0.10 \\
\hline
\end{tabular}
\end{table}

\section{The solar photospheric C abundance
\label{s:results}}

\subsection{The forbidden [C\,{\sc i}] 872.7\,nm line}

The forbidden [C\,{\sc i}] 872.7\,nm line has been studied in detail
by Allende Prieto et al. (2002) using the same 3D hydrodynamical
solar model atmosphere as employed here. They determined the
carbon abundance from a $\chi^2$-analysis of the solar flux atlas
of Kurucz et al. (1984) with three free parameters besides the
carbon abundance: the wavelength of the [C\,{\sc i}] line (see discussion
in Sect. \ref{s:atomicdata}), continuum placement and 
log\,$gf\cdot\epsilon_{\rm Si}$
for the Si\,{\sc i} 872.80\,nm line of which the [C\,{\sc i}] line
is located in its blue wing.
Including fitting errors and estimates of possible systematic errors,
they arrived at ${\rm log} \, \epsilon_{\rm C} = 8.39 \pm 0.04$.

For consistency with the analysis of the permitted lines, we have
re-analysed the forbidden line using the disk-center intensity profile
instead of the flux profile. The derived carbon abundance, however,
is not affected by this choice:
${\rm log} \, \epsilon_{\rm C} = 8.39 \pm 0.04$.
We note that Lambert \& Swings (1967) flagged a potential blending
Fe\,{\sc i} line, which, if contributing to the feature, would
imply a downward revision of our derived carbon abundances.
Using other Fe\,{\sc i} lines belonging to the same multiplet,
Allende Prieto et al. (2002) estimated the equivalent width of the line to
be $<0.03$\,pm (1\,pm $\equiv$ 10\,m\AA ), 
which corresponds to a downward change of $<0.02$\,dex
for the derived carbon abundance.

The best fit carbon abundance results in a theoretical disk-center
intensity equivalent width of 0.53\,pm. Employing this value, the
1D Holweger-M\"uller model atmosphere implies a carbon abundance of
${\rm log} \, \epsilon_{\rm C} = 8.45 \pm 0.04$ (adopting the same
error estimate as for the 3D analysis), while the {\sc marcs} model
suggests ${\rm log} \, \epsilon_{\rm C} = 8.40 \pm 0.04$. Clearly,
this particular transition is relatively insensitive to the choice of
model atmosphere.

\subsection{Permitted C\,{\sc i} lines}

\begin{table}[t!]
\caption{The derived solar carbon abundance as indicated by the
forbidden [C\,{\sc i}] 872.7\,nm transition and permitted C\,{\sc i} lines.
The results include the non-LTE abundance corrections presented
in Table \ref{t:nlte}. In the absence of detailed 3D non-LTE calculations
for C\,{\sc i}, the 1D non-LTE corrections for the {\sc marcs} model
atmosphere has here been used also for the 3D case.
\label{t:CI}
}
\begin{tabular}{lccccccc}
 \hline
line   & $\chi_{\rm exc}$ & log\,$gf$ & 3D$^{\rm a}$ &
$W_\lambda^{\rm b}$ & HM$^{\rm c}$ & {\sc marcs}$^{\rm c}$ \\
$$[nm] & [eV]             &           &              &
[pm]                &              &                       \\
 \hline
[C\,{\sc i}]: \\
 872.71 &    1.264 &   -8.136 &   8.39 &   0.53 &   8.45 &   8.40 \\
C\,{\sc i}: \\
 505.2167 &    7.685 &   -1.304 &   8.34 &   4.07 &   8.40 &   8.34 \\
 538.0337 &    7.685 &   -1.615 &   8.36 &   2.52 &   8.42 &   8.38 \\
 658.7610 &    8.537 &   -1.021 &   8.31 &   1.79 &   8.37 &   8.33 \\
 711.1469 &    8.640 &   -1.074 &   8.30 &   1.37 &   8.35 &   8.32 \\
 711.3179 &    8.647 &   -0.762 &   8.40 &   2.85 &   8.45 &   8.41 \\
 960.3036 &    7.480 &   -0.895 &   8.35 &   9.60 &   8.34 &   8.32 \\
1075.3976 &    7.488 &   -1.598 &   8.37 &   4.69 &   8.38 &   8.36 \\
1177.7546 &    8.643 &   -0.490 &   8.38 &   6.69 &   8.38 &   8.36 \\
1254.9493 &    8.847 &   -0.545 &   8.39 &   5.90 &   8.41 &   8.37 \\
1256.2124 &    8.848 &   -0.504 &   8.39 &   6.24 &   8.41 &   8.37 \\
1256.9042 &    8.848 &   -0.586 &   8.39 &   5.59 &   8.41 &   8.37 \\
1258.1585 &    8.848 &   -0.509 &   8.37 &   6.13 &   8.40 &   8.36 \\
2102.3151 &    9.172 &   -0.437 &   8.42 &   8.76 &   8.44 &   8.39 \\
2290.6565 &    9.172 &   -0.182 &   8.33 &  11.88 &   8.34 &   8.30 \\
3085.4621 &    9.736 &    0.086 &   8.35 &   5.53 &   8.36 &   8.34 \\
3406.5790 &    9.989 &    0.454 &   8.34 &   7.09 &   8.34 &   8.34 \\
\hline
\end{tabular}
\begin{list}{}{}
\item[$^{\rm a}$] The abundances derived using the 3D model atmosphere have
been obtained from profile fitting of the observed lines.
\item[$^{\rm b}$] The predicted disk-center intensity
equivalent widths with the 3D model atmosphere
using the best fit abundances shown in the fourth column.
\item[$^{\rm c}$] The derived abundances with the Holweger-M\"uller (1974)
and the {\sc marcs} (Asplund et al. 1997) 1D hydrostatic model atmospheres
in order to reproduce the equivalent widths presented in the fifth column.
\end{list}
\end{table}

\begin{figure*}[t]
\resizebox{\hsize}{!}{\includegraphics{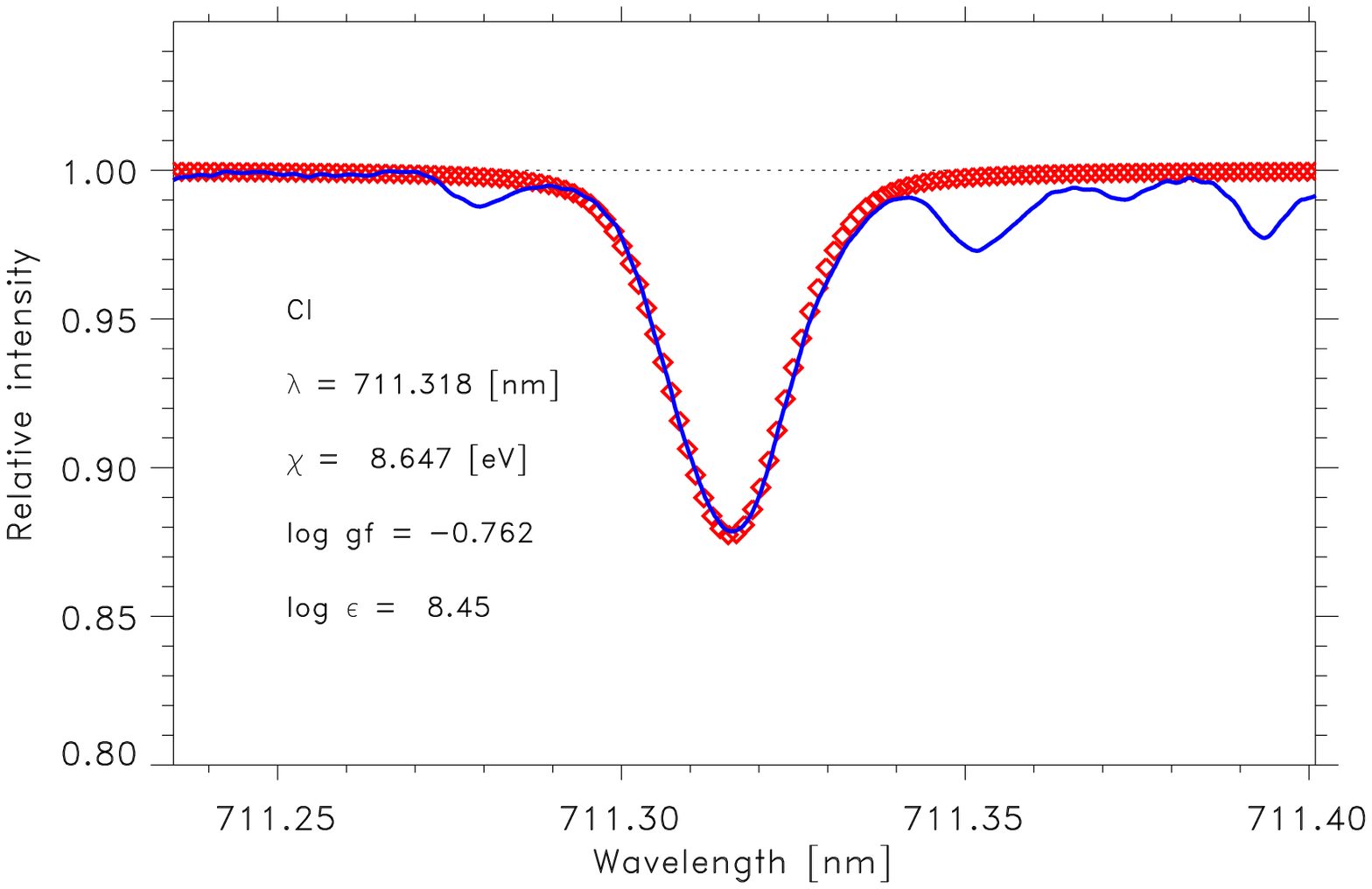}\includegraphics{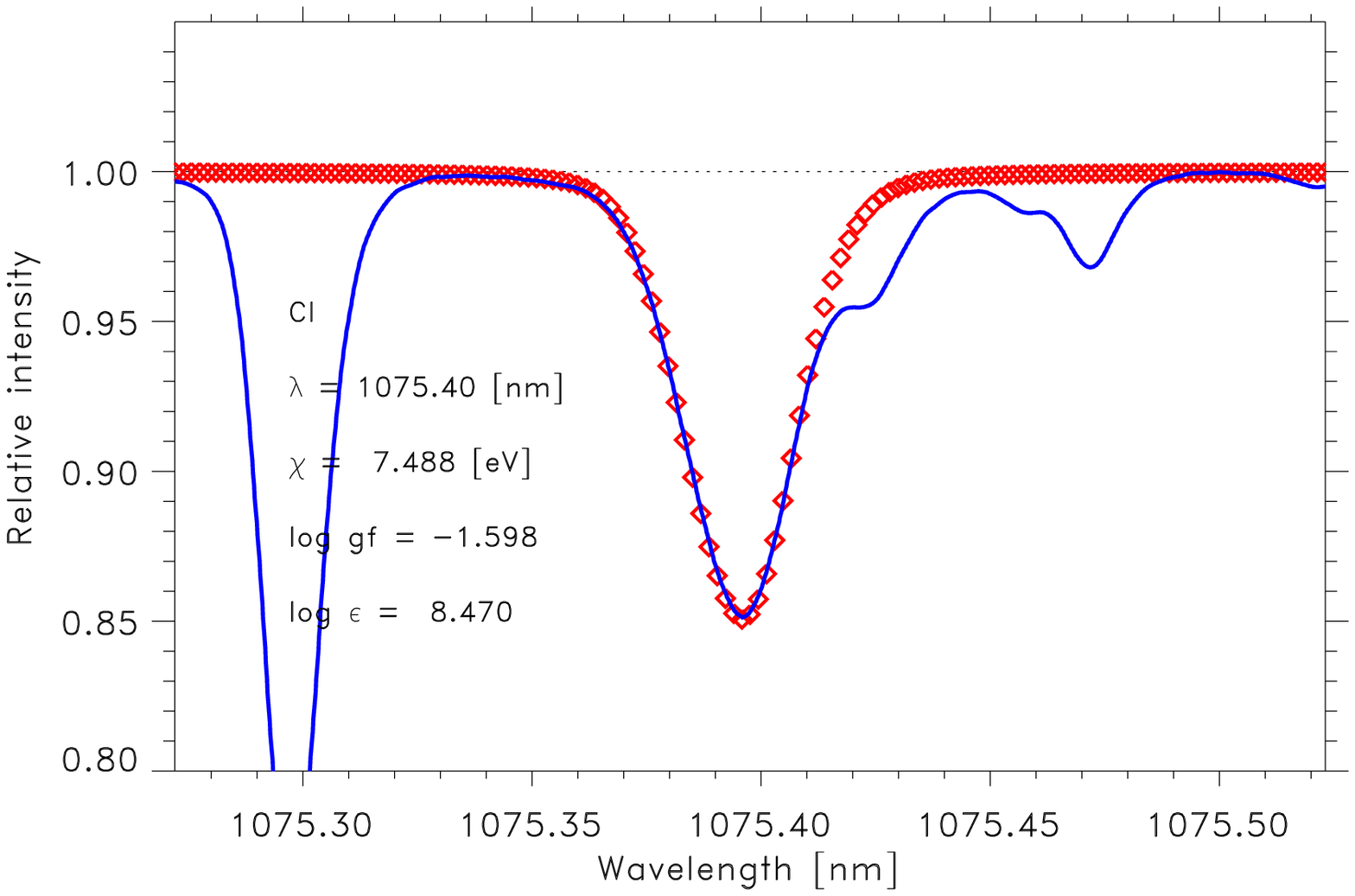}}
\resizebox{\hsize}{!}{\includegraphics{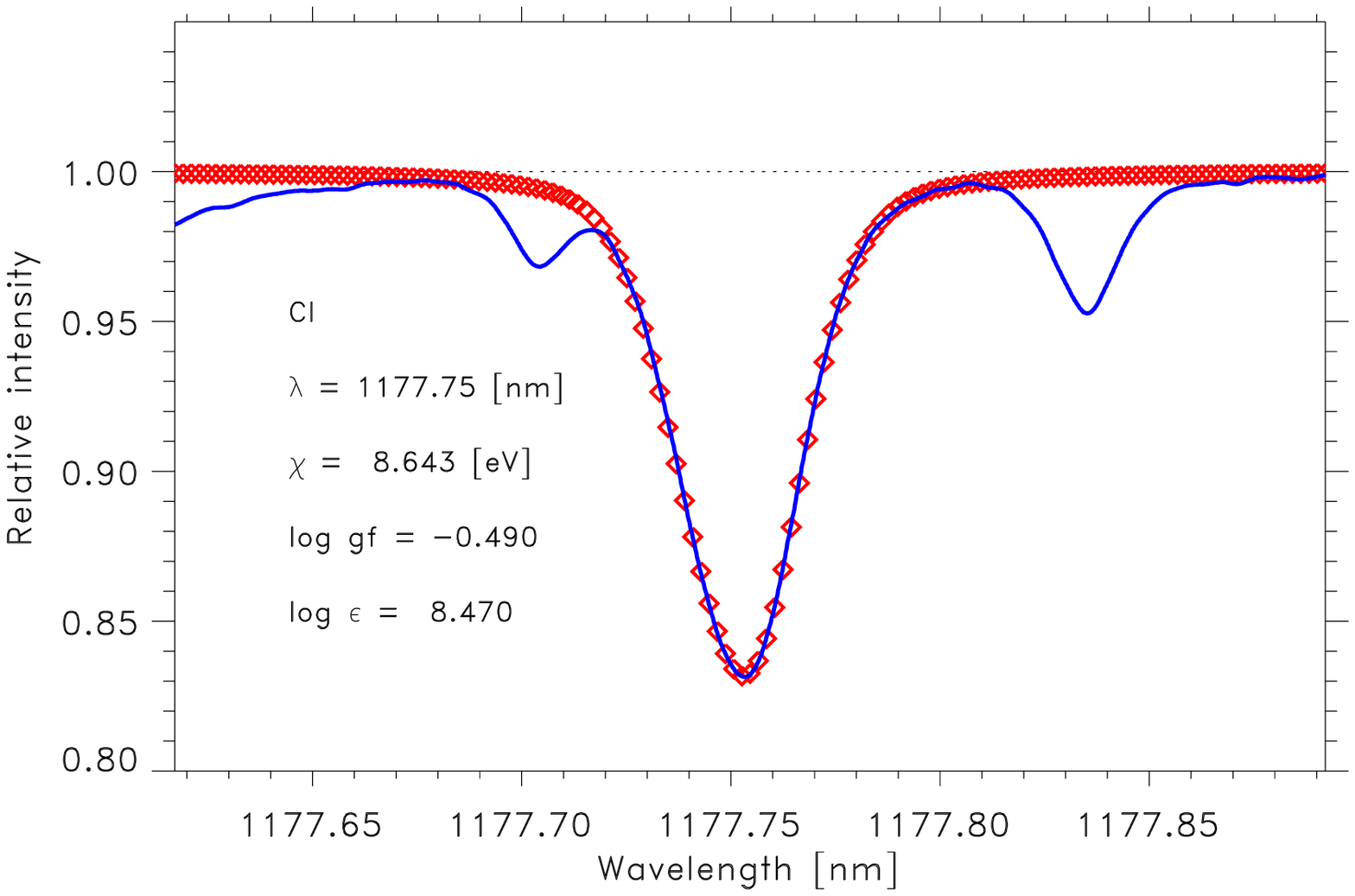}\includegraphics{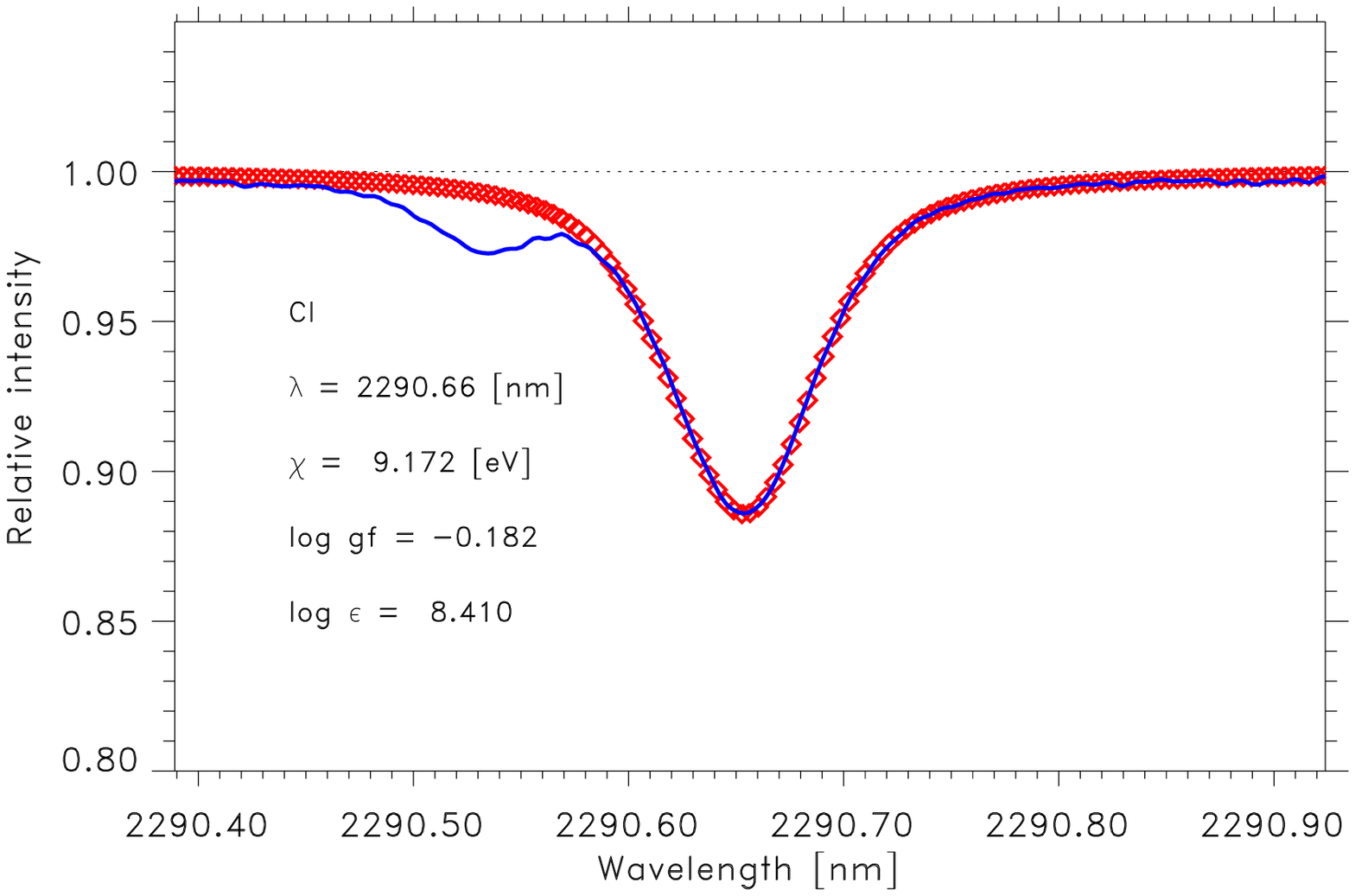}}
\caption{The predicted 3D LTE profiles of the
C\,{\sc i} 711.3, 1075.4, 1177.7 and 2290.6\,nm lines (diamonds) 
compared with the solar intensity atlas (solid lines, Brault \& Neckel 1987 for
the first three lines and Farmer \& Norton 1989 for the last line).
The excellent agreement is achieved without any micro- or macroturbulence
due to the Doppler shifts arising from the convective motions.
}
         \label{f:CIprof_3D}
\end{figure*}

The carbon abundance determination using permitted atomic lines
is based on the 16 C\,{\sc i} lines listed in Table \ref{t:CI}.
The individual abundances have been estimated using profile fitting
for disk-center intensity; we note for completeness
that very similar abundances would
have been derived had flux profiles been used instead.
The profile fitting has been performed for the spatially and
temporally averaged 3D LTE line calculations from the 100 snapshots used.
As clear from Fig. \ref{f:CIprof_3D} the agreement between predicted
and observed profiles is very satisfactory.
In contrast, the theoretical line profiles in 1D require additional
broadening from macroturbulence to get the correct line widths but 
of course the observed line shifts and asymmetries can not be accounted for.

Due to the large carbon model atom required to capture the
essence of the non-LTE effects, 3D non-LTE line formation computations
for the C\,{\sc i} lines have not been possible.
As a substitute, we apply the 1D non-LTE abundance corrections estimated
with the {\sc marcs} model atmosphere given in Table \ref{t:nlte} to
the 3D LTE abundance estimates to arrive at our final results presented in
Table \ref{t:CI}.
It is worthwhile pointing out that the similarities between
the 1D non-LTE results for the {\sc marcs} and Holweger-M\"uller
model atmospheres, in spite of their quite different temperature structures
(see Fig. 2 in Paper IV), give reasons to believe that the
3D non-LTE case would indeed be similar as well had it been available.

\begin{figure}[t]
\resizebox{\hsize}{!}{\includegraphics{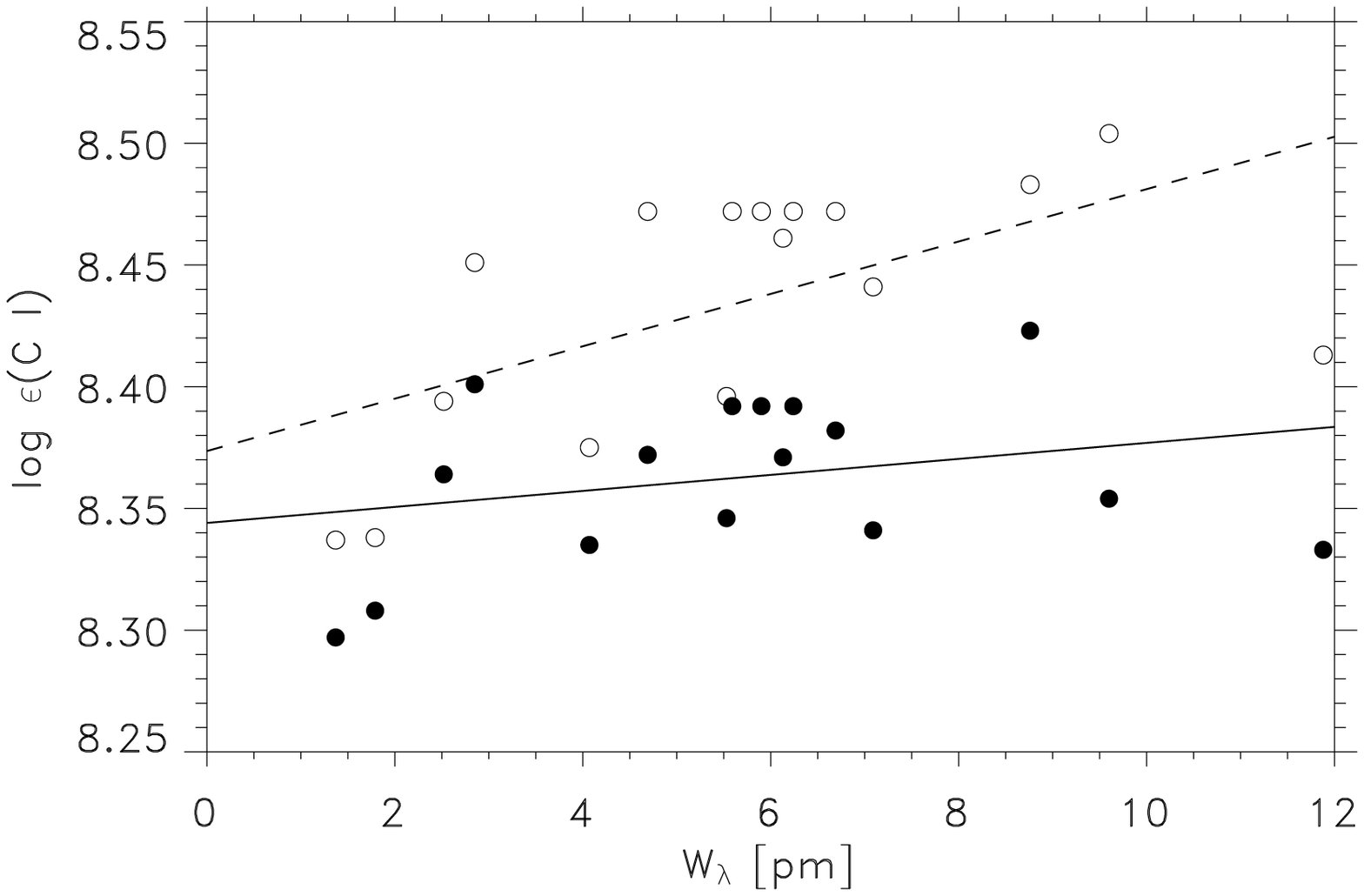}}
\resizebox{\hsize}{!}{\includegraphics{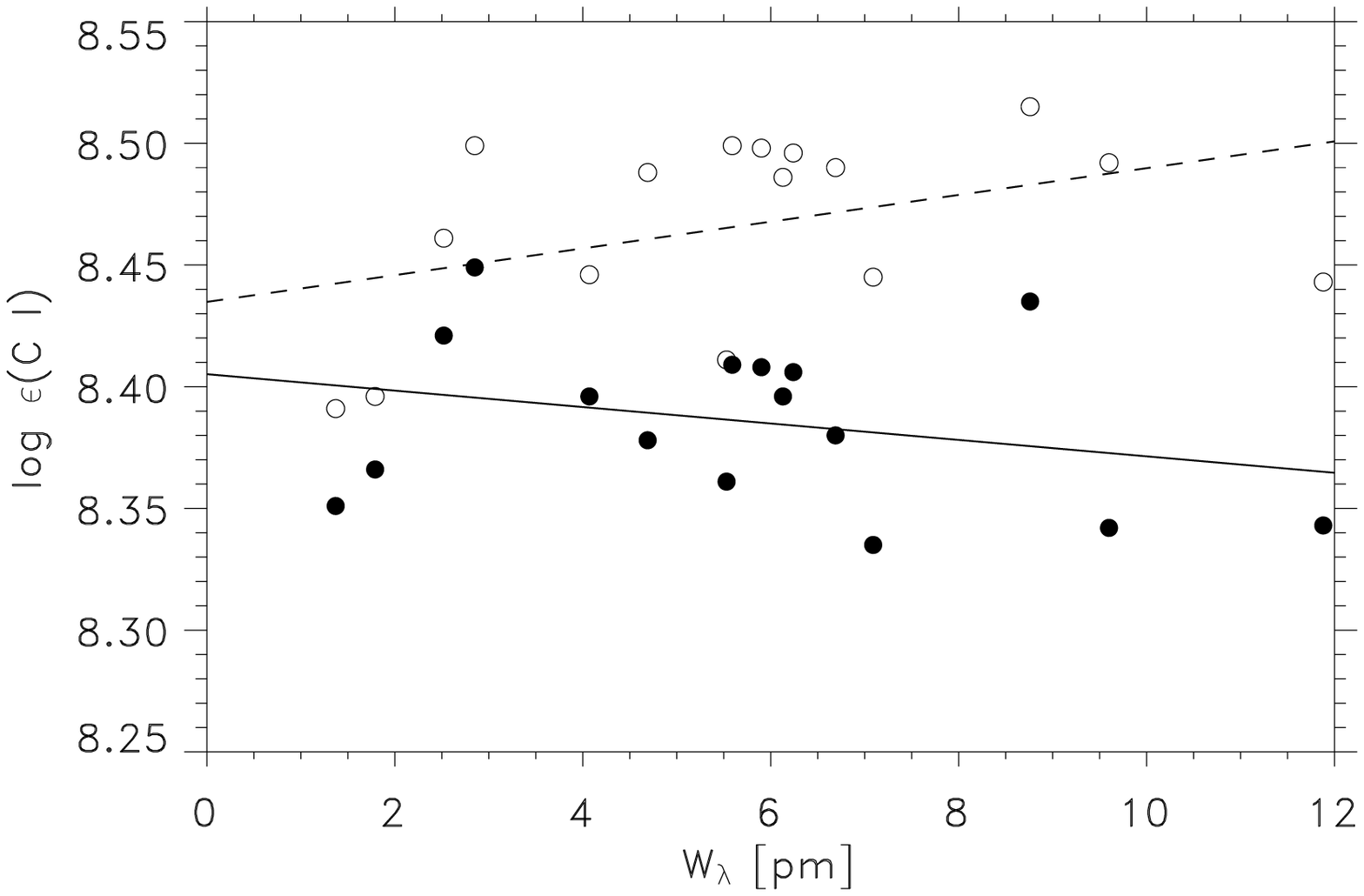}}
\resizebox{\hsize}{!}{\includegraphics{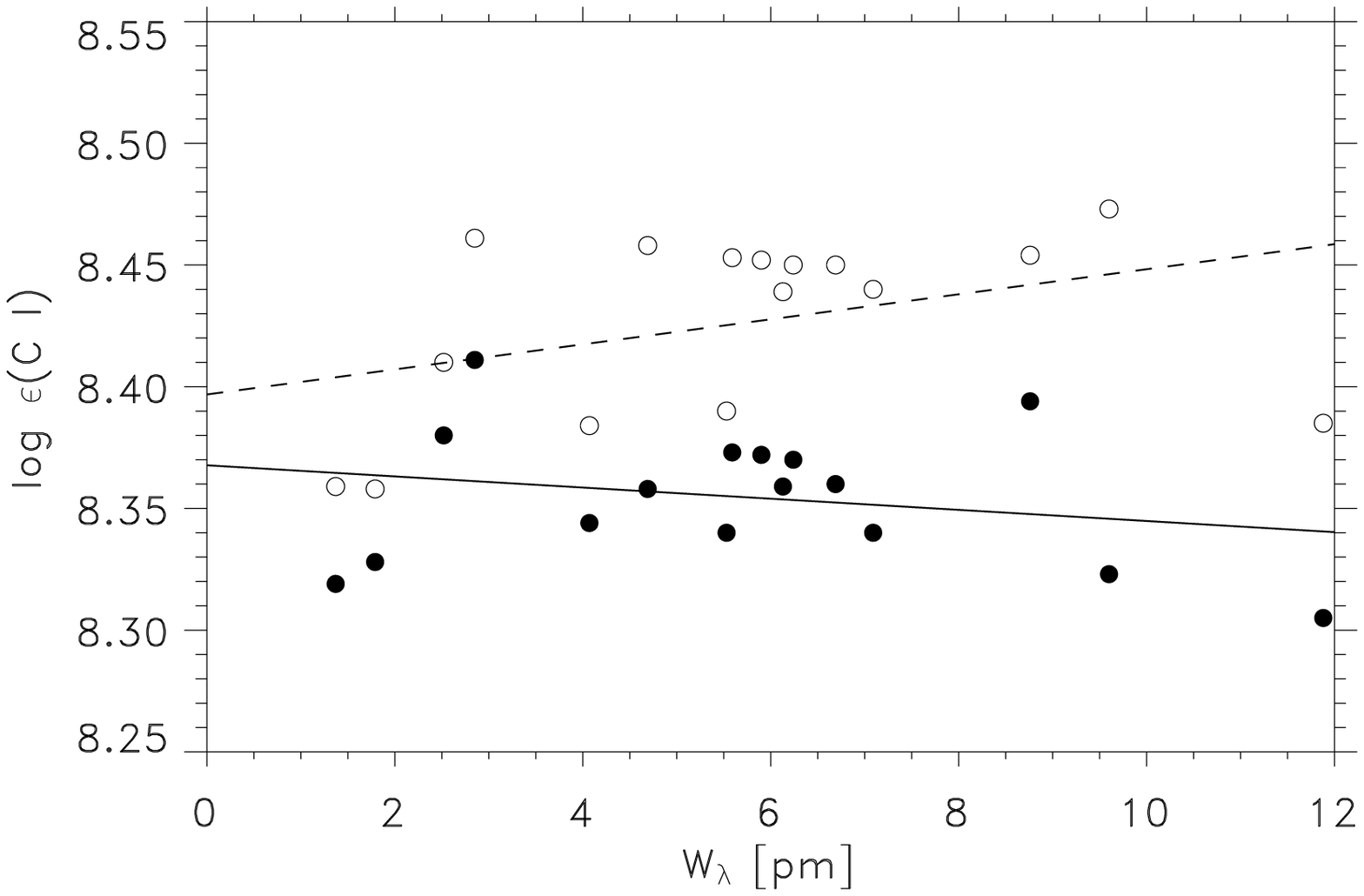}}
\caption{The derived solar carbon abundance from permitted C\,{\sc i}
lines using the 3D hydrodynamical model atmosphere as a function
of equivalent width ({\em Upper panel}). The filled circles denote the case when
the 1D {\sc marcs}-based non-LTE abundance corrections in
Table \ref{t:nlte} have been applied, while the open circles
correspond to the LTE results.
Also shown are the corresponding cases with the Holweger-M\"uller
({\em Middle panel}) and the {\sc marcs} ({\em Lower panel})
1D model atmospheres, which display similar trends with equivalent widths.
As explained in the text, this may signal underestimated non-LTE effects.
}
         \label{f:CI_Wlam}
\end{figure}

The mean C\,{\sc i}-based solar carbon non-LTE abundance is
${\rm log} \, \epsilon_{\rm C} = 8.36 \pm 0.03$ (standard deviation). 
The scatter is encouragingly small. 
There is no trend in the derived abundances with either wavelength
or excitation potential although as seen in Fig. \ref{f:CI_Wlam} 
there is a small correlation
with the strengths of the lines (about 0.04\,dex from the weakest to the 
strongest lines, i.e. $W_\lambda = 1-12$\,pm). 
One possibility is that the line broadening arising from
the Doppler shifts due to the convective motions are underestimated,
akin to using too small a microturbulence in 1D. Our previous 
3D LTE analysis of Fe\,{\sc i} lines have revealed a similar
trend (Asplund et al. 2000c). However it should be pointed out that
that correlation could also be the result of neglecting 
departures from LTE 
(Shchukina \& Trujillo Bueno 2001), and there is no
corresponding trend for Fe\,{\sc ii} or Si\,{\sc i} lines
(Asplund 2000), which would be expected if the convective
line broadening is underestimated. 
From Fig. \ref{f:CI_Wlam} it is clear that the
trend with equivalent width is largely 
driven by the two weakest lines, 658.7 and
711.1\,nm. Although the former line is located in the far wing of
the H$\alpha$ line and there are some telluric H$_2$O lines around
the latter, we have not been able to identify any obvious 
explanation for these two abundances being underestimated. 

In our opinion, the most likely explanation for this slight
correlation is underestimated non-LTE corrections in the 3D case.
We note that we have here adopted the non-LTE calculations from
the 1D {\sc marcs} model in the lack of suitable full 3D non-LTE
calculations. It is quite conceivable that the presence of
temperature inhomogeneities could produce slightly more pronounced
non-LTE effects, in particular for the stronger lines.
Fortunately, this residual trend is very small and will not
affect the mean abundance from the C\,{\sc i} lines 
significantly. 
Another possibility, while less likely, is that the $gf$-values 
for the C\,{\sc i} 658.7 and 711.1\,nm are over-estimated by about $0.05$\,dex.

We emphasize that the trend with equivalent width 
would have been more pronounced, as seen in Fig. \ref{f:CI_Wlam},
had departures from LTE not been considered (Fig. \ref{f:CI_Wlam}) 
and the line-to-line scatter would
have been significantly higher ($\pm 0.05$ instead of $\pm 0.03$).
This is a clear indication that these highly excited C\,{\sc i}
lines are indeed not formed in LTE. 
 
Using the equivalent widths estimated from the 3D profile fitting
and the available non-LTE corrections,
the 1D carbon abundances become
${\rm log} \, \epsilon_{\rm C} = 8.35 \pm 0.03$ with the {\sc marcs} and
${\rm log} \, \epsilon_{\rm C} = 8.39 \pm 0.03$ with the Holweger-M\"uller
model atmospheres. As for the 3D case, without application of the
computed non-LTE corrections there would be a pronounced trend with
equivalent widths (Fig. \ref{f:CI_Wlam}).
However, in both 1D cases the non-LTE 
abundances exhibit a trend 
in the opposite sense with respect to the LTE case.

\begin{figure}[t]
\resizebox{\hsize}{!}{\includegraphics{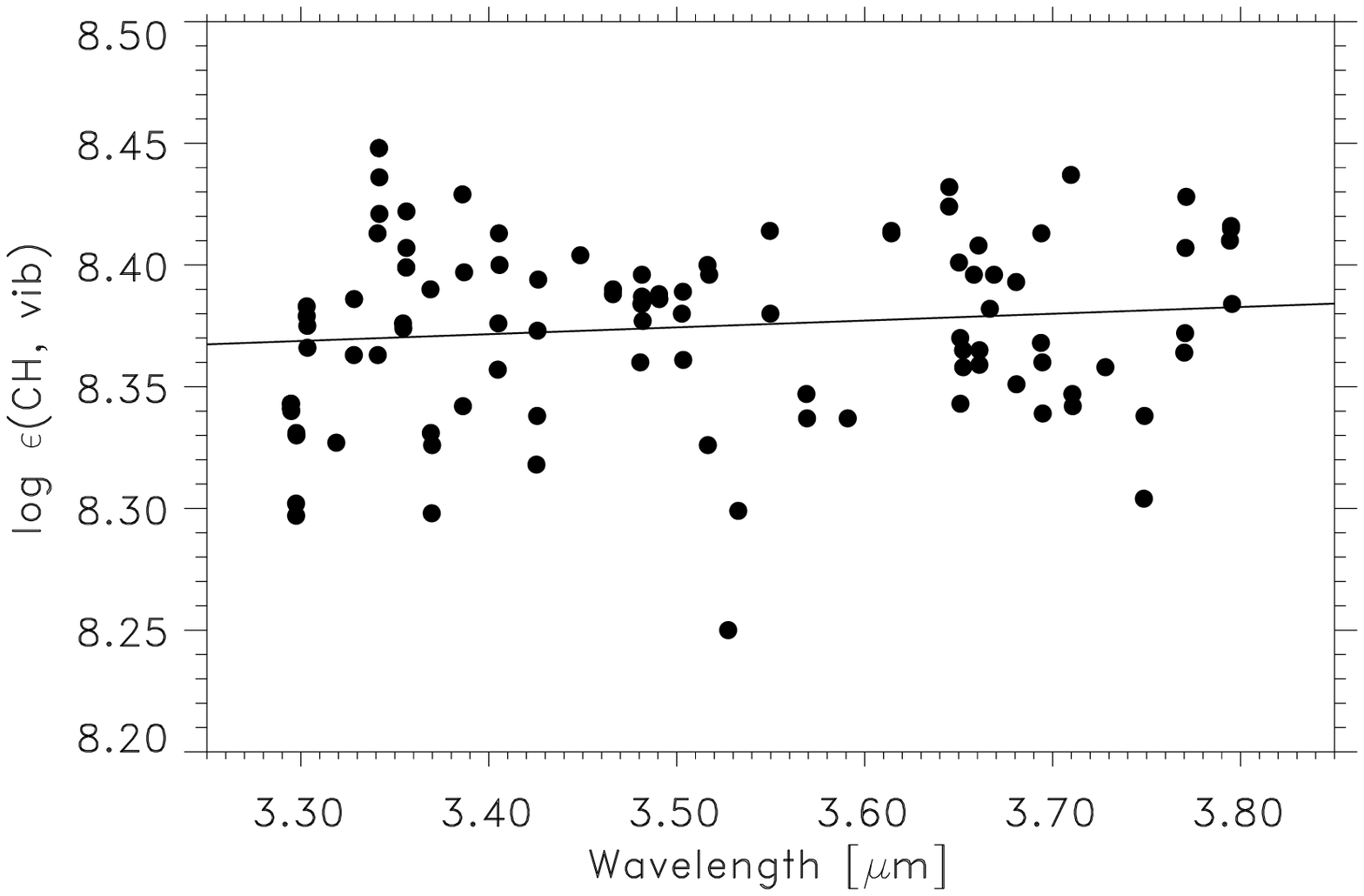}}
\resizebox{\hsize}{!}{\includegraphics{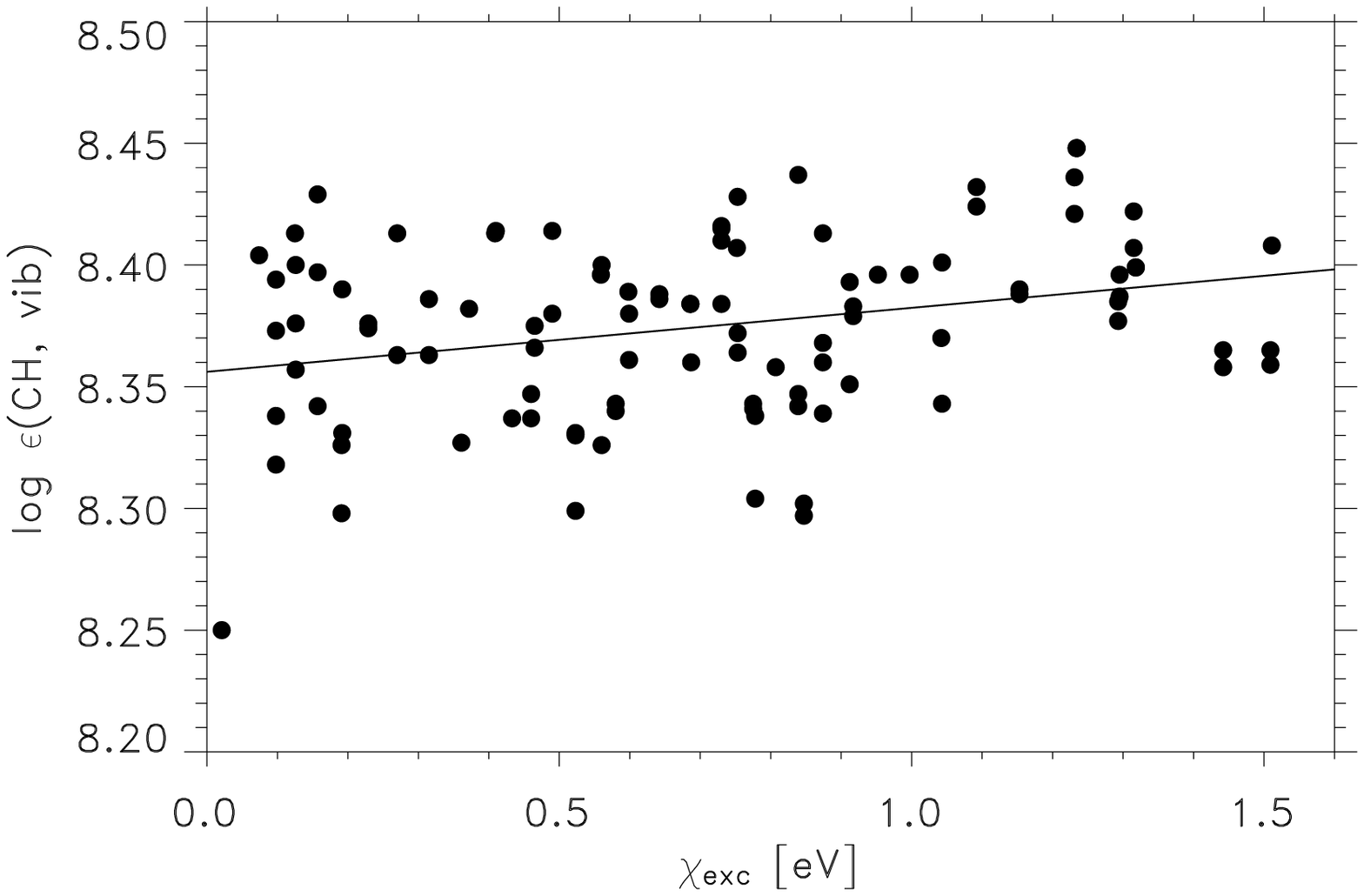}}
\resizebox{\hsize}{!}{\includegraphics{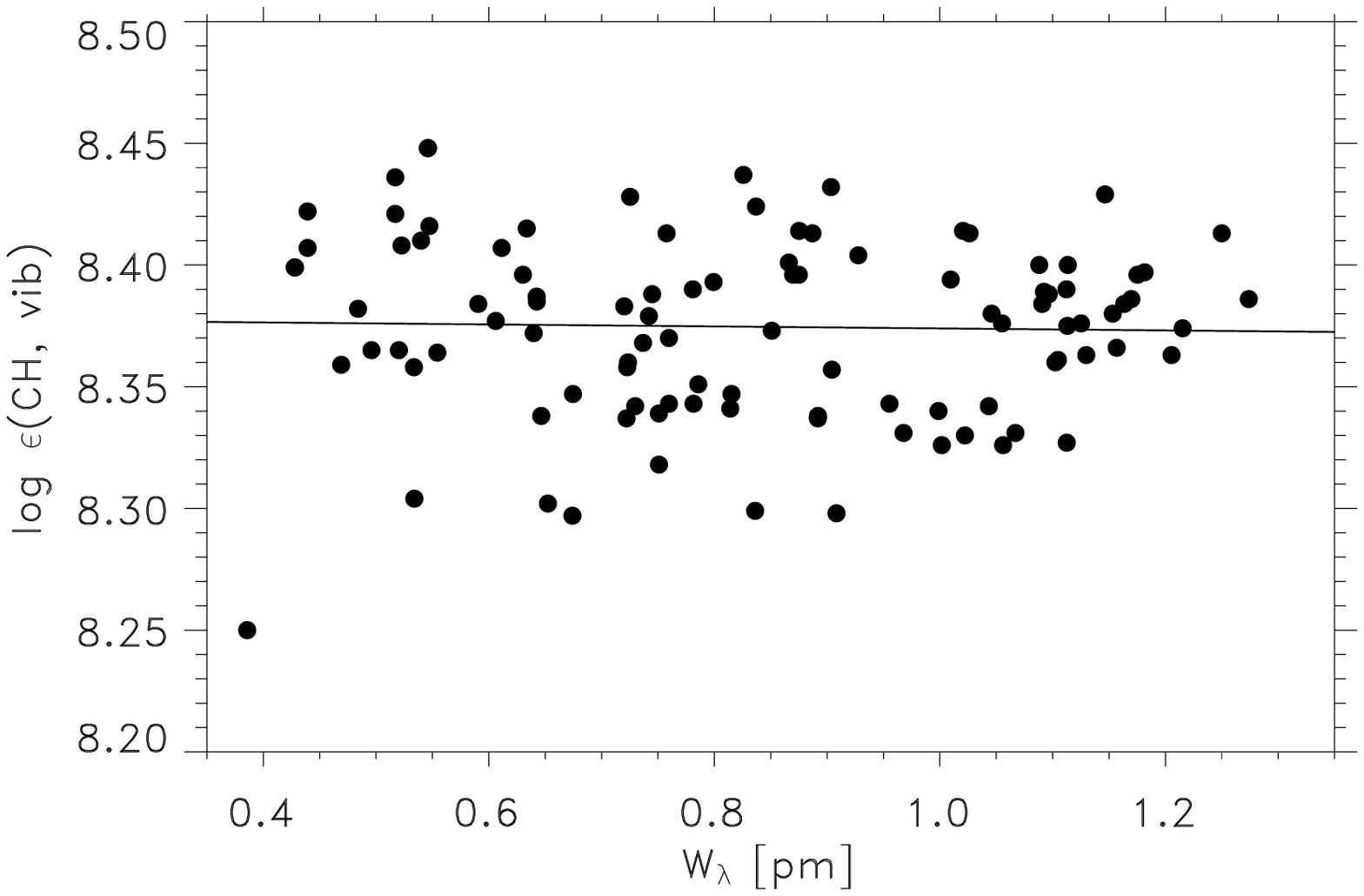}}
\caption{The derived solar carbon abundance (filled circles)
from CH vibration-rotation
lines using the 3D hydrodynamical
time-dependent simulation of the solar atmosphere (Asplund et al. 2000b)
as a function of wavelength, lower level excitation
potential and equivalent width (in pm).
The solid lines denote least-square-fits giving equal weights to all lines.
}
         \label{f:CHvr_3D}
\end{figure}

\subsection{CH vibration-rotation lines
\label{s:CHvr}}

High-quality, high resolution solar IR spectra like the
Spacelab-3 ATMOS atlas
open new opportunities for element abundance determinations
compared with the traditional UV-optical region employed in
most studies. 
One of the major advantages with the IR is the relatively clean
spectrum with few blending lines. In addition, many noteworthy
molecules have their vibration and rotation lines accessible 
in this region, which can provide a multitude of unperturbed
lines to base an abundance analysis on. 
Grevesse et al. (1991) reported the first solar carbon abundance
determination using CH vibration-rotation lines in the IR
and argued that these lines constitute one of the most reliable
sources for determinations of the solar carbon abundance.
Here we agree with this conclusion, but note that this is
only true when the lines are analysed with a realistic model of the
solar atmosphere. Due to the temperature sensitivity of
molecule formation in general (e.g. Asplund \& Garc\'{\i}a P{\'e}rez 2001), 
the strengths of
molecular transitions and consequently the derived
abundances depend crucially on the photospheric 
temperature structure. In addition, the presence of
atmospheric inhomogeneities 
induced by for example the convective motions can 
also strongly alter the equivalent widths of the lines. 
As these concerns are properly addressed by our use
of a realistic 3D hydrodynamical solar model atmosphere,
we consider the CH vibration-rotation lines as primary
abundance indicators.

\begin{table*}[t!]
\caption{The derived solar carbon abundance from CH vibration-rotation lines.
The individual abundances have been derived using measured disk-center ($\mu = 0.935$)
intensity equivalent widths for both the 3D model as well as the Holweger-M\"uller
and {\sc marcs} 1D model atmospheres. 
\label{t:CH_vr}
}
\begin{tabular}{lccccccclcccccccc}
 \hline
line   & $\chi_{\rm exc}$ & log\,$gf$ & $W_\lambda$ &
3D & HM & {\sc marcs} &&
line   & $\chi_{\rm exc}$ & log\,$gf$ & $W_\lambda$ &
3D & HM & {\sc marcs} \\
$$[nm] & [eV]   &   &  [pm]  &   &   &  &&
$$[nm] & [eV]   &   &  [pm]  &   &   &   \\
 \hline
 3294.5908 &    0.775 &   -2.910 &     0.78 &     8.34 &     8.49 &     8.38 &&
 3481.8975 &    1.293 &   -2.522 &     0.64 &     8.38 &     8.52 &     8.42 \\
 3294.6397 &    0.775 &   -2.890 &     0.81 &     8.34 &     8.49 &     8.38 &&
 3490.5767 &    0.642 &   -2.937 &     1.10 &     8.39 &     8.54 &     8.44 \\
 3294.8066 &    0.580 &   -3.016 &     0.96 &     8.34 &     8.50 &     8.39 &&
 3490.7277 &    0.642 &   -2.906 &     1.17 &     8.39 &     8.54 &     8.43 \\
 3294.8760 &    0.580 &   -2.993 &     1.00 &     8.34 &     8.50 &     8.38 &&
 3502.7022 &    0.599 &   -2.950 &     1.15 &     8.38 &     8.54 &     8.43 \\
 3297.4334 &    0.847 &   -2.877 &     0.65 &     8.30 &     8.45 &     8.34 &&
 3503.2787 &    0.598 &   -2.984 &     1.09 &     8.39 &     8.55 &     8.44 \\
 3297.4692 &    0.847 &   -2.858 &     0.67 &     8.30 &     8.45 &     8.34 &&
 3503.4344 &    0.599 &   -2.950 &     1.10 &     8.36 &     8.52 &     8.41 \\
 3297.5795 &    0.523 &   -3.055 &     0.97 &     8.33 &     8.49 &     8.38 &&
 3516.3324 &    0.560 &   -3.035 &     1.09 &     8.40 &     8.56 &     8.45 \\
 3297.6587 &    0.523 &   -3.030 &     1.02 &     8.33 &     8.49 &     8.38 &&
 3516.5314 &    0.560 &   -2.998 &     1.00 &     8.33 &     8.48 &     8.38 \\
 3303.0757 &    0.917 &   -2.845 &     0.72 &     8.38 &     8.53 &     8.42 &&
 3517.2491 &    0.559 &   -2.998 &     1.18 &     8.40 &     8.55 &     8.45 \\
 3303.1132 &    0.917 &   -2.827 &     0.74 &     8.38 &     8.53 &     8.41 &&
 3527.3702 &    0.021 &   -3.890 &     0.39 &     8.25 &     8.43 &     8.31 \\
 3303.4083 &    0.465 &   -3.095 &     1.11 &     8.37 &     8.54 &     8.42 &&
 3532.7151 &    0.523 &   -3.089 &     0.84 &     8.30 &     8.46 &     8.35 \\
 3303.4954 &    0.465 &   -3.069 &     1.16 &     8.37 &     8.53 &     8.41 &&
 3549.5444 &    0.490 &   -3.148 &     1.02 &     8.41 &     8.57 &     8.47 \\
 3318.8332 &    0.361 &   -3.152 &     1.11 &     8.33 &     8.49 &     8.38 &&
 3549.8184 &    0.490 &   -3.103 &     1.05 &     8.38 &     8.54 &     8.43 \\
 3328.1821 &    0.315 &   -3.228 &     1.13 &     8.36 &     8.53 &     8.42 &&
 3568.9890 &    0.460 &   -3.212 &     0.82 &     8.35 &     8.51 &     8.40 \\
 3328.3163 &    0.315 &   -3.197 &     1.27 &     8.39 &     8.55 &     8.44 &&
 3569.3191 &    0.460 &   -3.162 &     0.89 &     8.34 &     8.50 &     8.39 \\
 3340.7684 &    0.270 &   -3.278 &     1.25 &     8.41 &     8.58 &     8.47 &&
 3590.9704 &    0.433 &   -3.283 &     0.72 &     8.34 &     8.50 &     8.39 \\
 3340.9122 &    0.270 &   -3.244 &     1.21 &     8.36 &     8.53 &     8.42 &&
 3614.1794 &    0.410 &   -3.297 &     0.88 &     8.41 &     8.58 &     8.47 \\
 3341.5359 &    1.234 &   -2.716 &     0.55 &     8.45 &     8.59 &     8.48 &&
 3614.2293 &    0.409 &   -3.362 &     0.76 &     8.41 &     8.58 &     8.47 \\
 3341.5359 &    1.234 &   -2.716 &     0.55 &     8.45 &     8.59 &     8.48 &&
 3644.9021 &    1.092 &   -2.641 &     0.84 &     8.42 &     8.56 &     8.46 \\
 3341.7395 &    1.231 &   -2.732 &     0.52 &     8.44 &     8.56 &     8.45 &&
 3645.0087 &    1.092 &   -2.615 &     0.90 &     8.43 &     8.57 &     8.47 \\
 3341.7557 &    1.231 &   -2.716 &     0.52 &     8.42 &     8.56 &     8.45 &&
 3650.1702 &    1.043 &   -2.652 &     0.87 &     8.40 &     8.54 &     8.44 \\
 3354.3361 &    0.229 &   -3.330 &     1.13 &     8.38 &     8.55 &     8.43 &&
 3650.8169 &    1.042 &   -2.680 &     0.76 &     8.37 &     8.51 &     8.41 \\
 3354.5025 &    0.229 &   -3.293 &     1.22 &     8.37 &     8.54 &     8.43 &&
 3650.9366 &    1.043 &   -2.652 &     0.76 &     8.34 &     8.48 &     8.38 \\
 3356.0419 &    1.318 &   -2.691 &     0.43 &     8.40 &     8.56 &     8.45 &&
 3652.3483 &    1.442 &   -2.444 &     0.52 &     8.36 &     8.49 &     8.40 \\
 3356.0419 &    1.318 &   -2.691 &     0.43 &     8.40 &     8.56 &     8.45 &&
 3652.4003 &    1.442 &   -2.425 &     0.53 &     8.36 &     8.49 &     8.39 \\
 3356.1651 &    1.315 &   -2.706 &     0.44 &     8.42 &     8.53 &     8.43 &&
 3658.0986 &    0.997 &   -2.691 &     0.87 &     8.40 &     8.54 &     8.44 \\
 3356.1760 &    1.315 &   -2.691 &     0.44 &     8.41 &     8.54 &     8.44 &&
 3660.5436 &    1.511 &   -2.416 &     0.52 &     8.41 &     8.54 &     8.44 \\
 3368.9374 &    0.192 &   -3.386 &     1.11 &     8.39 &     8.56 &     8.45 &&
 3661.0114 &    1.509 &   -2.416 &     0.47 &     8.36 &     8.49 &     8.39 \\
 3369.1470 &    0.192 &   -3.346 &     1.07 &     8.33 &     8.50 &     8.39 &&
 3661.0562 &    1.509 &   -2.398 &     0.50 &     8.36 &     8.49 &     8.39 \\
 3369.6174 &    0.191 &   -3.386 &     0.91 &     8.30 &     8.47 &     8.36 &&
 3666.5972 &    0.372 &   -3.566 &     0.48 &     8.38 &     8.55 &     8.44 \\
 3369.8123 &    0.191 &   -3.346 &     1.06 &     8.33 &     8.50 &     8.38 &&
 3668.8128 &    0.952 &   -2.733 &     0.87 &     8.40 &     8.54 &     8.44 \\
 3385.9743 &    0.157 &   -3.447 &     1.15 &     8.43 &     8.60 &     8.49 &&
 3680.5876 &    0.912 &   -2.810 &     0.80 &     8.39 &     8.54 &     8.44 \\
 3386.2232 &    0.157 &   -3.402 &     1.04 &     8.34 &     8.52 &     8.40 &&
 3680.7623 &    0.912 &   -2.776 &     0.79 &     8.35 &     8.50 &     8.39 \\
 3386.8611 &    0.157 &   -3.402 &     1.18 &     8.40 &     8.57 &     8.46 &&
 3693.7566 &    0.874 &   -2.859 &     0.74 &     8.37 &     8.52 &     8.41 \\
 3404.7668 &    0.126 &   -3.513 &     0.90 &     8.36 &     8.53 &     8.42 &&
 3693.9761 &    0.874 &   -2.822 &     0.89 &     8.41 &     8.56 &     8.46 \\
 3405.0678 &    0.126 &   -3.463 &     1.06 &     8.38 &     8.55 &     8.44 &&
 3694.5273 &    0.874 &   -2.859 &     0.72 &     8.36 &     8.51 &     8.40 \\
 3405.3892 &    0.125 &   -3.513 &     1.03 &     8.41 &     8.59 &     8.47 &&
 3694.7288 &    0.874 &   -2.822 &     0.75 &     8.34 &     8.49 &     8.38 \\
 3405.6716 &    0.126 &   -3.463 &     1.11 &     8.40 &     8.57 &     8.46 &&
 3709.7385 &    0.839 &   -2.912 &     0.83 &     8.44 &     8.59 &     8.48 \\
 3425.3356 &    0.098 &   -3.585 &     0.75 &     8.32 &     8.49 &     8.38 &&
 3710.4902 &    0.839 &   -2.912 &     0.67 &     8.35 &     8.49 &     8.39 \\
 3425.7085 &    0.098 &   -3.529 &     0.89 &     8.34 &     8.51 &     8.40 &&
 3710.7255 &    0.839 &   -2.872 &     0.73 &     8.34 &     8.49 &     8.39 \\
 3425.9194 &    0.098 &   -3.585 &     0.85 &     8.37 &     8.55 &     8.43 &&
 3728.0673 &    0.807 &   -2.925 &     0.72 &     8.36 &     8.51 &     8.41 \\
 3426.2715 &    0.098 &   -3.529 &     1.01 &     8.39 &     8.57 &     8.46 &&
 3748.5501 &    0.778 &   -3.033 &     0.53 &     8.30 &     8.45 &     8.35 \\
 3448.6936 &    0.074 &   -3.602 &     0.93 &     8.40 &     8.58 &     8.47 &&
 3748.8874 &    0.778 &   -2.983 &     0.65 &     8.34 &     8.49 &     8.39 \\
 3466.0327 &    1.153 &   -2.599 &     0.74 &     8.39 &     8.53 &     8.42 &&
 3770.0378 &    0.753 &   -3.102 &     0.55 &     8.36 &     8.52 &     8.41 \\
 3466.0773 &    1.153 &   -2.580 &     0.78 &     8.39 &     8.53 &     8.42 &&
 3770.4823 &    0.753 &   -3.046 &     0.64 &     8.37 &     8.52 &     8.42 \\
 3480.5974 &    0.687 &   -2.862 &     1.10 &     8.36 &     8.52 &     8.41 &&
 3770.6875 &    0.752 &   -3.102 &     0.61 &     8.41 &     8.56 &     8.46 \\
 3481.2203 &    0.686 &   -2.892 &     1.09 &     8.38 &     8.54 &     8.43 &&
 3771.1057 &    0.753 &   -3.046 &     0.73 &     8.43 &     8.58 &     8.48 \\
 3481.3393 &    0.686 &   -2.863 &     1.16 &     8.38 &     8.54 &     8.43 &&
 3794.3253 &    0.730 &   -3.182 &     0.54 &     8.41 &     8.56 &     8.46 \\
 3481.4645 &    1.295 &   -2.539 &     0.63 &     8.40 &     8.53 &     8.43 &&
 3794.8896 &    0.730 &   -3.116 &     0.63 &     8.41 &     8.57 &     8.47 \\
 3481.4931 &    1.295 &   -2.522 &     0.64 &     8.39 &     8.52 &     8.42 &&
 3794.9254 &    0.730 &   -3.182 &     0.55 &     8.42 &     8.57 &     8.47 \\
 3481.8682 &    1.293 &   -2.539 &     0.61 &     8.38 &     8.51 &     8.41 &&
 3795.4612 &    0.730 &   -3.116 &     0.59 &     8.38 &     8.54 &     8.43 \\
\hline
\end{tabular}
\end{table*}

In this study we rely on 102 CH vibration-rotation lines
of the (1,0), (2,1) and (3,2) bands around 3.3-3.8\,$\mu$m, 
for which the equivalent widths were measured in the 
Spacelab-3 ATMOS IR intensity ($\mu=0.935$, i.e. nearly
at solar disk-center) atlas. 
With our 3D model and assuming LTE for the line formation
and molecular concentration, these CH lines imply 
a carbon abundance of 
${\rm log} \, \epsilon_{\rm C} = 8.38 \pm 0.04$.
The derived abundances for each line together with the employed 
line data are listed in Table \ref{t:CH_vr}.
As seen in Fig. \ref{f:CHvr_3D}, there are no trends in
derived carbon abundance with wavelength or equivalent width,
although there is a weak correlation with lower excitation
potential ($0.03\pm0.01$\,dex/eV). It should be noted however that
the total span in excitation potential is only 1.4\,eV.
We do not consider this weak trend as flagging a serious
problem, in particular given the encouragingly small scatter.
These lines are also formed in a rather narrow region of
the atmosphere (mean optical depth of line formation
$-1.25 \la {\rm log} \tau_{\rm 500} \la -1.15$ in the
Holweger-M\"uller model atmosphere; line formation
depth is a ill-defined concept in a 3D model atmosphere).
We have not been able to identify any particular problem
with the CH 3527.3\,nm line, which gives a significantly
lower C abundance (${\rm log} \, \epsilon_{\rm C} = 8.25$) 
than the other lines. 

The corresponding result using the 
Holweger-M\"uller model atmosphere gives a substantially 
higher result:
${\rm log} \, \epsilon_{\rm C} = 8.53 \pm 0.04$.
This is a direct consequence of the higher temperatures
in this 1D model. 
In this case the abundance trend with excitation potential
has disappeared and instead been replaced with a minor
correlation with equivalent width.
This opposite behaviour 
compared with the 3D case reflects the correlation
between equivalent width and excitation potential of the
employed CH lines. 
With the {\sc marcs} model atmosphere, the CH lines
indicate a carbon abundance of 
${\rm log} \, \epsilon_{\rm C} = 8.42 \pm 0.04$
without any apparent trends with transition properties.

\subsection{C$_2$ electronic lines}

There are numerous weak lines from the C$_2$ Swan band in the solar spectrum.
From these we selected a subsample of 17 (0,0) lines between 495 and 516\,nm,
which are all apparently undisturbed by neighboring lines.
The lines employed here have equivalent widths between 0.3 and 1.4\,pm,
which make the derived abundances insensitive to the velocity broadening.
Since the previous uncertainties surrounding the C$_2$ dissociation energy
appear to have dissolved (see Sect. \ref{s:atomicdata}),
we include these lines among our primary abundance indicators.

With the 3D hydrodynamical solar model atmosphere the measured
equivalent widths of these 17 C$_2$ lines
yield a solar carbon abundance of
${\rm log} \, \epsilon_{\rm C} = 8.44 \pm 0.03$.
There are no significant abundance trends with line properties
according to the results presented in Table \ref{t:C2_Swan}.
The C$_2$ based abundance is in excellent agreement with the
values derived from the other preferred diagnostics.
The corresponding results for the 1D models are
${\rm log} \, \epsilon_{\rm C} = 8.53 \pm 0.03$ (Holweger-M\"uller) and
${\rm log} \, \epsilon_{\rm C} = 8.46 \pm 0.03$ ({\sc marcs}).

\begin{table}[t!]
\caption{The derived solar carbon abundance as indicated by lines from
the C$_2$ (0,0) Swan band. As for the other carbon abundance diagnostics
the abundances have been derived for the 3D model and the Holweger-M\"uller
and {\sc marcs} model atmospheres. The measured equivalent widths are for
disk center intensity ($\mu = 1.0$).
\label{t:C2_Swan}
}
\begin{tabular}{lccccccc}
 \hline
line   & $\chi_{\rm exc}$ & log\,$gf$ & $W_\lambda$ &
3D & HM & {\sc marcs} \\
$$[nm] & [eV]             &           &       [pm]        &                &
             &                       \\
 \hline
  495.1400 &    1.070 &    0.631 &     0.50 &     8.46 &     8.55 &     8.47 \\
  499.2300 &    0.820 &    0.572 &     0.65 &     8.43 &     8.51 &     8.44 \\
  503.3700 &    0.570 &    0.181 &     0.46 &     8.42 &     8.51 &     8.43 \\
  503.7700 &    0.580 &    0.486 &     1.15 &     8.48 &     8.57 &     8.50 \\
  505.2600 &    0.500 &    0.450 &     1.00 &     8.42 &     8.51 &     8.44 \\
  507.3600 &    0.370 &    0.073 &     0.56 &     8.41 &     8.50 &     8.43 \\
  508.6200 &    0.340 &    0.345 &     1.35 &     8.47 &     8.56 &     8.49 \\
  510.3700 &    0.260 &    0.267 &     1.25 &     8.45 &     8.54 &     8.47 \\
  510.9100 &    0.230 &    0.237 &     1.25 &     8.45 &     8.54 &     8.47 \\
  510.9300 &    0.220 &   -0.088 &     0.54 &     8.41 &     8.50 &     8.43 \\
  513.2500 &    0.140 &   -0.243 &     0.49 &     8.42 &     8.52 &     8.44 \\
  513.6600 &    0.110 &   -0.343 &     0.53 &     8.47 &     8.57 &     8.50 \\
  514.0400 &    0.100 &   -0.405 &     0.37 &     8.42 &     8.52 &     8.44 \\
  514.3300 &    0.100 &   -0.414 &     0.46 &     8.47 &     8.57 &     8.49 \\
  514.4900 &    0.100 &   -0.450 &     0.34 &     8.42 &     8.52 &     8.44 \\
  515.0500 &    0.340 &    0.332 &     1.03 &     8.41 &     8.50 &     8.43 \\
  516.3400 &    0.170 &    0.113 &     0.93 &     8.40 &     8.50 &     8.43 \\
\hline
\end{tabular}
\end{table}

\subsection{CH electronic lines}

While there is a swath of CH electronic lines in the optical solar
spectrum the vast majority are badly blended with other lines or are
too strong to yield accurate results. We have identified only nine
apparently unblended lines from the (0,0) and (1,1) bands of CH A-X
between 421 and 436\,nm.
Their equivalent widths are between 3.5 and 8\,pm, making them partly
saturated and thus sensitive to the atmospheric velocity field.
That together with the rather crowded spectral region the lines
are located in, relegates
the CH electronic lines to carry only a secondary role in our
solar C abundance analysis.

The measured equivalent widths of these nine CH lines together
with the 3D hydrodynamical model atmosphere yield a
C abundance of ${\rm log} \, \epsilon_{\rm C} = 8.45 \pm 0.04$.
No significant trends with wavelength, excitation potential or
equivalent width are present (Table \ref{t:CH_AX}), 
although the spans in these parameters
are quite small.
The derived 3D abundance is slightly higher 
than those from the primary abundance indicators (C\,{\sc i}, [C\,{\sc i}],
CH vibration and C$_2$ electronic lines).
The corresponding results with the Holweger-M\"uller and
{\sc marcs} model atmospheres are
${\rm log} \, \epsilon_{\rm C} = 8.59 \pm 0.04$ and
${\rm log} \, \epsilon_{\rm C} = 8.44 \pm 0.04$, respectively.
As for the CH vibration lines, the higher temperatures in the
Holweger-M\"uller model automatically lead to higher abundances.

\begin{table}[t!]
\caption{The derived solar carbon abundance based on CH A-X electronic
lines in the optical using three different model atmospheres. The
listed equivalent widths are measured in the disk-center ($\mu =1.0$) 
intensity solar atlas.
\label{t:CH_AX}
}
\begin{tabular}{lccccccc}
 \hline
line   & $\chi_{\rm exc}$ & log\,$gf$ & $W_\lambda$ &
3D & HM & {\sc marcs} \\
$$[nm] & [eV]             &           &       [pm]        &                &
             &                       \\
 \hline
  421.8723 &    0.413 &   -1.008 &     7.88 &     8.41 &     8.55 &     8.40 \\
  424.8945 &    0.192 &   -1.423 &     6.52 &     8.42 &     8.56 &     8.41 \\
  425.3003 &    0.523 &   -1.523 &     3.70 &     8.44 &     8.57 &     8.43 \\
  425.3209 &    0.523 &   -1.486 &     4.00 &     8.45 &     8.58 &     8.44 \\
  425.5252 &    0.157 &   -1.453 &     6.80 &     8.45 &     8.59 &     8.44 \\
  426.3976 &    0.460 &   -1.593 &     3.90 &     8.48 &     8.61 &     8.47 \\
  427.4186 &    0.074 &   -1.558 &     6.59 &     8.45 &     8.59 &     8.44 \\
  435.6375 &    0.157 &   -1.830 &     4.74 &     8.54 &     8.68 &     8.53 \\
  435.6600 &    0.157 &   -1.775 &     4.20 &     8.40 &     8.54 &     8.40 \\
\hline
\end{tabular}
\end{table}

\subsection{Summary}

Of the five carbon abundance indicators considered here,
the greatest weight is given to the forbidden [C\,{\sc i}], 
permitted C\,{\sc i}, CH vibration-rotation 
and C$_2$ electronic lines, with the CH electronic lines only
assigned a supporting role in view of their broadening
sensitivity and location in a relatively crowded spectral region. 
The four primary abundance indicators imply highly concordant
results when analysed with the 3D hydrodynamical model atmosphere,
as summarised in Table \ref{t:abund}. 
This is particularly noteworthy given the
very different temperature sensitivity and
distinct formation depths of these lines. 
In sharp contrast, the Holweger-M\"uller model atmosphere gives much more
disparate results, with the molecular lines 
suggesting much higher abundances than the atomic transitions
as a consequence of the different temperature structures and
lack of photospheric inhomogeneities. 
The theoretical {\sc marcs} model, however, performs nearly
as well as the 3D model in this respect; a similar conclusion
would likely have held had we used for example a Kurucz theoretical
1D model atmosphere. 

The errors quoted in Table \ref{t:abund} for the different
types of transitions reflect the line-to-line scatter 
rather than smaller standard deviation of the mean, except for
the forbidden [C\,{\sc i}] 872.7\,nm which also includes 
estimates for fitting errors and systematic errors. 
The mean of the primary indicators becomes
${\rm log} \, \epsilon_{\rm C} = 8.39 \pm 0.03$.
The dominant source of error though is likely to be of
systematic nature. The excellent agreement between the different types
of lines gives us some confidence that unknown systematic errors
can not be very significant and estimate them to be on
the order of $\pm 0.05$ for the mean abundance.
We therefore arrive at our best estimate of the solar carbon
abundance as:

$${\rm log} \, \epsilon_{\rm C} = 8.39 \pm 0.05.$$

\begin{table}[t!]
\caption{The derived solar carbon abundance as indicated by
forbidden [C\,{\sc i}], permitted C\,{\sc i}, CH vibration-rotation,
C$_2$ electronic and CH electronic lines. The results
for the C\,{\sc i} lines include non-LTE abundance corrections.
The quoted uncertainties only reflect the line-to-line
scatter for the different types of C diagnostics, except for
[C\,{\sc i}] where the error estimate includes uncertainties in
the profile fitting.
\label{t:abund}
}
\begin{tabular}{lccc}
 \hline
lines  & \multicolumn{3}{c}{${\rm log} \epsilon_{\rm C}$} \\
\cline{2-4}
         &   3D & HM & {\sc marcs} \\
 \hline
primary indicators: & & & \\
$$[C\,{\sc i}] & $8.39 \pm 0.04$ & $8.45 \pm 0.04$ & $8.40 \pm 0.04$ \\
C\,{\sc i}   & $8.36 \pm 0.03$ & $8.39 \pm 0.03$ & $8.35 \pm 0.03$ \\
CH vib-rot   & $8.38 \pm 0.04$ & $8.53 \pm 0.04$ & $8.42 \pm 0.04$ \\
C$_2$ electronic& $8.44 \pm 0.03$ & $8.53 \pm 0.03$ & $8.46 \pm 0.03$ \\
\hline
secondary indicator: & & & \\
CH electronic& $8.45 \pm 0.04$ & $8.59 \pm 0.04$ & $8.44 \pm 0.04$ \\
\hline
\end{tabular}
\end{table}

\section{Comparison with previous studies}

The solar carbon abundance derived in Sect. \ref{s:results} is
significantly lower than most previous estimates. In his extensive and
thorough survey of the solar C, N and O abundances,
Lambert (1978) found ${\rm log} \, \epsilon_{\rm C} = 8.69 \pm 0.10$
based on a combination of [C\,{\sc i}], CH A-X and C$_2$ lines using
the Holweger-M\"uller semi-empirical model atmosphere.
He preferred to exclude the permitted
C\,{\sc i} lines from the final average due to uncertainties stemming from
the available transition probabilities and possible departures from LTE, which
in the time since then have at least partly been addressed.
Likewise, the CH B-X, CH C-X and CO vibration lines, neither of which have
been considered here, were relegated to a supporting role in Lambert's study
for a variety of reasons. He did not have access to the CH vibration
lines which we rank as a primary abundance indicator.
In contrast, we consider the CH A-X lines as inferior to the
other indicators.
The main differences with Lambert's value for [C\,{\sc i}]
compared with our own re-analysis using the Holweger-M\"uller model originate
in the adopted $gf$-values ($-0.08$\,dex) and preferred disk-center
equivalent width ($-0.10$\,dex) but we have not been able to trace
the remaining $-0.06$\,dex discrepancy.
The use of a 3D model atmosphere instead of the Holweger-M\"uller model
further decreases the derived carbon abundance by $0.06$\,dex.

Lambert's carefully derived C abundance was subsequently duly
adopted by Grevesse (1984) in his review of the solar abundances.
Soon thereafter, Sauval \& Grevesse (1985) identified CH vibration lines in
the solar IR spectrum, which should enable a reliable abundance determination,
as also demonstrated herein.
In their widely used survey of the solar and meteoritic abundances,
Anders \& Grevesse (1989) preferred ${\rm log} \, \epsilon_{\rm C} = 8.56 \pm 0.04$
based primarily on a preliminary analysis of the CH vibration lines using
Holweger-M\"uller model atmosphere.
Once the C analysis was complete, this value had been slightly revised to
${\rm log} \, \epsilon_{\rm C} = 8.60 \pm 0.05$
based on 104 CH vibration lines (Grevesse et al. 1991).
In addition, Grevesse et al.
analysed C$_2$ (Swan and Phillips bands), CH A-X and C\,{\sc i} lines and estimated
a mean solar carbon abundance of ${\rm log} \, \epsilon_{\rm C} = 8.60 \pm 0.05$.
The main reason for the
high molecular based abundances compared with ours clearly stems from
the use of the Holweger-M\"uller model instead of a 3D hydrodynamical model atmosphere,
which takes into account temperature inhomogeneities aand has a cooler
mean temperature stratification (Table \ref{t:abund}).
The difference with their C\,{\sc i} abundances is primarily a combination of them
neglecting non-LTE effects, choice of $gf$-values and their adopted equivalent widths,
with a minor part probably coming from the employed pressure broadening data.

More recently, Grevesse \& Sauval (1999) performed a re-analysis of the
C\,{\sc i}, [C\,{\sc i}], CH and C$_2$ lines. Based on the existence
of significant trends in the obtained Fe abundances with excitation potential
with the standard Holweger-M\"uller model atmosphere,
Grevesse et al. (1999) derived a new temperature structure which was
slightly cooler in the higher layers. 
This modified Holweger-M\"uller model atmosphere
forms the basis for the slightly lower solar C abundance estimated
by Grevesse \& Sauval (1998): ${\rm log} \, \epsilon_{\rm C} = 8.52 \pm 0.06$.
As before, the neglect of photospheric inhomogeneities causes the
molecular lines still to yield a too high abundance.

As described above, Allende Prieto et al. (2002) performed a detailed
study of the [C\,{\sc i}] using the same 3D hydrodynamical solar model
atmosphere as employed in the present work and found a much lower
value than other studies: ${\rm log} \, \epsilon_{\rm C} = 8.39 \pm 0.04$.
This low value is corroborated here by our more extensive
calculations also involving permitted C\,{\sc i} and molecular lines.
In contrast, Holweger (2001) concluded that the solar carbon abundance is
${\rm log} \, \epsilon_{\rm C} = 8.59 \pm 0.11$ based on C\,{\sc i} and
[C\,{\sc i}] lines when applying non-LTE ($-0.05$\,dex)
and granulation corrections ($+0.02$\,dex) to the 1D LTE results of
St\"urenburg \& Holweger (1990) and Bi\'emont et al. (1993).
We note that these granulation corrections have
the opposite sign to those estimated here (Table \ref{t:CI}) as
a result of Holweger's different definition of these (Paper IV).
The main differences to our
Holweger-M\"uller-based results lie in the selection of lines (we restricted
our list largely to lines with $W_\lambda \la 10$\,pm while Holweger's sample
contains a large number of lines with $10 \la W_\lambda \la 20$\,pm),
choice of $gf$-values, adopted equivalent widths and the inclusion or
not of inelastic H collisions in the non-LTE calculations.
Holweger did not consider molecular lines.

In summary, we are confident that our comprehensive 3D analysis
superseeds previous works devoted to the solar photospheric carbon abundance.
This conclusion is supported among other things
by the excellent agreement between the
different abundance indicators illustrated in Table \ref{t:abund}.
We emphasize that the transition from a 1D hydrostatic to a 3D hydrodynamical
model atmosphere is only part of the explanation for the large
downward revision of the solar carbon abundance compared with
for example Lambert (1978) and Grevesse et al. (1991).
Improved atomic and molecular data, higher quality
observations, accounting for non-LTE effects and identification
of existing blends also played important roles.

\section{Conclusions}

We have presented a comprehensive study of the available atomic and
molecular lines to derive a new solar carbon abundance which is
significantly lower than most previous analyses:
${\rm log} \, \epsilon_{\rm C} = 8.39 \pm 0.05$
(C/H\,$= 245 \pm 30 \cdot 10^{-6}$).
The quoted uncertainty includes our best estimate of the
remaining systematic errors.
Our primary abundance indicators are the forbidden [C\,{\sc i}] 872.7\,nm,
high-excitation permitted C\,{\sc i}, CH vibration-rotation 
and C$_2$ Swan lines
with a supporting role played by CH A-X electronic lines.
A special effort has been made to select the most accurate and reliable
atomic and molecular data such as transition probabilities and dissociation energies.
For the C\,{\sc i} lines detailed non-LTE calculations in a 1D context have been performed,
which revealed significant departures from LTE in the line formation.
Finally, a novel feature of the present study is the use of a realistic
3D hydrodynamical solar model atmosphere, which furthermore has proved to be crucial
in order to obtain consistent abundances between the various diagnostics.
In sharp contrast, the 1D Holweger-M\"uller model atmosphere
yields much higher abundances for the molecular lines than for the atomic lines,
partly due to the higher temperatures in the line-forming regions and partly
due to the neglect of temperature inhomogeneities.
The excellent concordance between the various transitions with widely
different temperature and pressure sensitivities is a very strong argument
in favour of both the 3D model atmosphere as such and the new low carbon
abundance.

Further support for the new low solar carbon abundance advocated herein
comes from a comparison of the chemical composition in related
environments. The revised solar photospheric carbon abundance is now
in good agreement with those measured in nearby B stars with
solar iron abundances: ${\rm log} \, \epsilon_{\rm C} = 8.28 \pm 0.17$
or C/H\,$= 190 \pm 90 \cdot 10^{-6}$
(Sofia \& Meyer 2001). Furthermore, the solar carbon abundance is now
in good agreement with estimates of the total (gas plus dust) local
interstellar medium carbon abundance for realistic gas-to-dust ratios
(Andr\'e et al. 2003; Estiban et al. 2004). Finally, coupled with the previously
re-determined solar photospheric oxygen abundance (Paper IV), the
photospheric C/O ratio is $0.54 \pm 0.09$, which is in good agreement with
the measurements in solar flares ($0.42 \pm 0.09$, Fludra et al. 1999;
$0.54 \pm 0.04$, Murphy et al. 1997) and solar wind particles
($0.47 \pm 0.01$, Reames 1999)\footnote{Since both carbon and
oxygen are high ionization potential elements, to first order it
is expected that the solar coronal C/O ratio is unaffected by
the so-called first ionization potential (FIP) effect, which
introduces a differential depletion of elements with low ionization potential
($\chi_{\rm ion} \la 9$\,eV) for as yet unexplained reasons, see review
by Raymond (1999).}.

The new solar carbon abundance corresponds to a change of $-0.28$\,dex 
relative to that advocated in Lambert (1978) and $-0.17$\,dex compared with
the preferred value in the standard source of Anders \& Grevesse (1989).
This change is primarily the outcome of improved models of the solar photosphere
and the line formation processes, as well as more accurate atomic
and molecular line data.
The possibility of similar substantial 
systematic errors in standard 1D abundance analyses
still today is certainly worth keeping in mind when interpreting the results
of other abundance studies 
in terms of stellar nucleosynthesis and galactic evolution.

\begin{acknowledgements}
We wish to thank Mats Carlsson,
Remo Collet, Damian Fabbian,
Ana Elia Garc\'{\i}a P{\'e}rez, Dan Kiselman, David Lambert,
\AA ke Nordlund, Bob Stein and Regner Trampedach for
discussions and expert assistance with model atmosphere and
line formation calculations.
We also thank the referee Jo Bruls for valuable comments. 
MA has been supported by research grants from
the Swedish Natural Science Foundation, the Royal Swedish Academy
of Sciences, the G\"oran Gustafsson Foundation, and the Australian Research Council.
CAP gratefully acknowledges support from NSF (AST-0086321) and
NASA (ADP02-0032-0106 and LTSA02-0017-0093).
NG appreciates financial support from the Royal Observatory (Brussels).
\end{acknowledgements}

\end{document}